\documentclass[pra,aps,11pt]{revtex4-2}
\usepackage{amsmath,amssymb,amsthm,amsfonts,bm,bbm,graphicx,times,psfrag}
\usepackage{physics}
\usepackage[utf8]{inputenc}
\usepackage[english]{babel}
\usepackage[pdftex]{color}
\usepackage[colorlinks, linkcolor=blue, citecolor=blue, urlcolor=blue, breaklinks=true]{hyperref}
\usepackage{microtype}
\usepackage{color}
\usepackage{soul}

\newcommand{\be} {\begin{equation}}
\newcommand{\ee} {\end{equation}}

\expandafter\let\csname equation*\endcsname\relax
\expandafter\let\csname endequation*\endcsname\relax
\usepackage{amsmath, times}
\usepackage{amssymb}
\usepackage{amsfonts}
\usepackage{mathrsfs}
\usepackage{physics}
\usepackage{mathbbol} 
\usepackage{forest}
\usepackage{tikz}
\usepackage{multirow}
\usetikzlibrary{positioning}

\usepackage{graphicx} 
\usepackage{physics}
\usepackage{amsmath}
\usepackage{bbold}
\usepackage{xcolor}
\usepackage{tikz}
\usetikzlibrary{arrows.meta} 
\usepackage{graphicx}
\usepackage{mathrsfs}

\begin{document}

\title[Quantum walks as a tool to design robust quantum batteries]{Quantum walks as a tool to design robust quantum batteries: the role of topology and chirality}

\author{Simone Cavazzoni, Giovanni Ragazzi}
\affiliation{Dipartimento di Scienze Fisiche, Informatiche e Matematiche,  Universit\`{a} di Modena e Reggio Emilia, I-41125 Modena, Italy}
\email{simone.cavazzoni@unimore.it, giovanni.ragazzi@unimore.it}
\author{Paolo Bordone}
\affiliation{Dipartimento di Scienze Fisiche, Informatiche e Matematiche,  Universit\`{a} di Modena e Reggio Emilia, I-41125 Modena, Italy}
\affiliation{Centro S3, CNR-Istituto di Nanoscienze, I-41125 Modena, Italy}
\email{paolo.bordone@unimore.it}
\author{Matteo G. A. Paris}
\affiliation{Dipartimento di Fisica {\em Aldo Pontremoli}, Universit\`{a} di Milano, 
I-20133 Milano, Italy}
\affiliation{Istituto Nazionale di Fisica Nucleare, Sezione di Milano,  
I-20133 Milano, Italy}
\email{matteo.paris@fisica.unimi.it}
\date{\today}
\begin{abstract}
The maximum work that can be extracted from a quantum battery is bounded by the ergotropy of the system, which is determined by the spectral properties of the Hamiltonian. In this paper, we employ the formalism of quantum walks to investigate how the topology of the battery and the chirality of the Hamiltonian influence its performance as an energy storage unit. We analyze architectures of battery cells based on ring, complete, and wheel graph structures and analyze their behavior in the presence of noise. Our results show that these structures exhibit distinct ergotropy scaling, with the interplay between chirality and topology providing a tunable mechanism to optimize work extraction and enhance robustness against decoherence. In particular, chirality enhances ergotropy for complete quantum cells, without altering the linear scaling with size, whereas in ring cells, it bridges the performance gap between configurations with odd and even number of units. Additionaly, chirality may be exploited to force degeneracies in the Hamiltonian, a condition that can spare the ergotropy to vanish in the presence of pure dephasing. We conclude that topology and chirality are key resources for improving ergotropy, offering guidelines to optimize quantum energy devices and protocols.
\end{abstract}
\maketitle
\section{Introduction}
\label{Sec:Intro}
The miniaturization of devices, and the potential advantages offered by quantum 
systems, have driven significant interest in developing temporary quantum-based 
energy storage devices, usually referred to as \textit{quantum batteries} \cite{alicki2013entanglement,campaioli2024colloquium}. These systems operate by transitioning between two fundamental states: an \textit{active} state, from which work can be extracted, and a \textit{passive} state, which cannot yield any further work under unitary operations. This state is uniquely determined by the Hamiltonian of the system, up to spectral degeneracies \cite{binder2015quantacell,crescente2024boosting}. The maximum work extractable from such an isolated quantum system is termed \textit{ergotropy}\cite{allahverdyan2004maximal}, a key figure of merit that defines the thermodynamic efficiency of the battery. Once the system reaches a passive state, it must be recharged, i.e., returned to an active state, to enable further energy extraction. Thus, the two essential protocols governing quantum batteries are work extraction and recharging. In the case of closed systems, governed by unitary operations, these processes are effectively symmetric, differing only in their initial and final states. The development of general and efficient protocols for these operations remains a central challenge in quantum thermodynamics.
Beyond theoretical proposals, experimental realizations of quantum batteries have been explored across various platforms, including quantum processors \cite{gemme2022ibm}, artificial atom arrays \cite{liu2019loss}, spin chains \cite{le2018spin}, and quantum dots \cite{maillette2023experimental,de2023coherence}. Notably, solid-state architectures may show collective advantage in charging power \cite{ferraro2018high,Maciej20,quach2022superabsorption}, making them particularly promising for practical implementations. Recent advances have extended 
these demonstrations to organic systems, spins, and superconducting circuits \cite{quach2022superabsorption,joshi2022experimental,hu2022optimal,santos2019stable,dou2022highly}. At the same time, research continues to explore how quantum effects, especially coherence, can be harnessed to enhance battery performance \cite{crescente2020ultrafast,PhysRevLett.125.180603,ghosh2020enhancement,ghosh2021fast}, paving the way for more efficient and powerful energy storage technologies.

Understanding the structural properties of quantum batteries \cite{PhysRevA.110.012227,Camposeo25} is crucial for optimizing their performance, as their energy storage and extraction capabilities are deeply influenced by the underlying physical architecture. The arrangement of energy levels, coupling mechanisms, and topology within the battery directly impact its ergotropy, charging protocols, and resilience to decoherence \cite{Maciej20}. To systematically explore these effects, a versatile and inherently quantum mechanical framework is required, one that naturally incorporates discrete spatial and energetic degrees of freedom \cite{PhysRevE.71.036128}. Quantum walks provide precisely such a framework, offering a unified description of coherent state transfer, excitation dynamics, and energy redistribution in discrete systems \cite{mulken2011continuous}. By modeling quantum batteries within the quantum walk formalism, one can investigate how structural properties, such as connectivity, local potentials, and symmetry constraints, affect the thermodynamical features of the system and enable the design of tailored architectures for enhanced energy storage and retrieval. In addition to topology, we explore the role of \textit{chirality}, i.e., the presence of complex phases in the off-diagonal elements of the Hamiltonian. Those phases break the symmetry under the time-reversal transformation $t \rightarrow -t$, making chirality a resource for quantum state transfer \cite{cavazzoni2025perfect}, transport efficiency ~\cite{cavazzoni2022perturbed}, and spatial routing protocols ~\cite{bottarelli2023quantum,ragazzi2025scalable}. While these studies primarily focus on position-space dynamics, we here investigate how chirality influences energy extraction and storage in \textit{momentum space}, providing new insights into the design of quantum batteries with directional control over work protocols.

In the following, we first discuss general properties of energy extraction in closed and open systems to highlight the role of the Hamiltonian spectrum and illustrate the degrees of freedom at hand. We then introduce three different quantum battery models based on chiral quantum cells, termed \textit{Q-Cells}, and discuss their ergotropy, charging protocols, and resilience to decoherence.  Our results show that the three architectures exhibit distinct 
ergotropy scaling: among ring cells, those with even number of sites achieve maximal work extraction, while odd rings require chiral phases to bridge this gap. By using ring cells one has ergotropy independent on the size $N$ of the system, while complete cells outperform the others with ergotropy growing linearly with size, and wheel Q-Cells show a $\sqrt{N}$ scaling. Despite this scaling, however, the practical challenges in constructing fully connected batteries must be taken into account. This consideration, combined with the size-independent ergotropy of ring-shaped batteries, led us to investigate the behavior of ring-based Q-Cells with 3 and 4 vertices more closely, particularly analyzing their performance under noise and decoherence.

The paper is thus structured as follows. In Section~\ref{sec:QWEiaN}, we analyze the theory of quantum work extraction for both pure and mixed quantum states. Section~\ref{sec:FQB} examines various noise sources that degrade performance, offering insights into the resilience and efficiency of quantum batteries under non-ideal conditions. In Section~\ref{sec:IpfWE}, we derive a general form of unitary work extraction and recharge protocols applicable to pure and mixed states.   Section~\ref{sec:QWT} introduces the theoretical framework of quantum walks, including the definition of chirality and the mathematical formalism for describing quantum batteries. Sections~\ref{sec:ALV}--\ref{sec:TCDC} present a prototypical quantum battery model, along with its general charge and discharge protocols. Building on this, Sections~\ref{sec:SB}--\ref{sec:CB} propose distinct designs for chiral Q-Cells, each accompanied by their respective unitary work extraction and recharge transformations. 
In Section \ref{sec:NC} we analyze the performance of noisy ring Q-Cells and in
Section~\ref{sec:CaEE} we discuss the genuinely quantum advantages enabled by chirality 
in our battery models. Section \ref{sec:C} closes the paper with some concluding remarks.

\section{Quantum work extraction in a nutshell}
\label{sec:QWEiaN}
The second law of thermodynamics provides fundamental limitations to energy transformations in thermodynamical processes. In particular, these limitations set the upper limits for power and efficiency. If no external driving is present, the system thermalyzes by decreasing its free energy. For a microscopic system isolated from its environment, the relevant quantity is no longer the free energy, but rather the so called \textit{ergotropy} \cite{allahverdyan2004maximal} or its conditional counterpart \cite{francica2017daemonic,PhysRevApplied.20.044073}.
\subsection{Ergotropy for pure states}
For a $N$ dimensional finite-size quantum system described by the Hamiltonian $\mathcal{H}$
\begin{equation}
    \mathcal{H} = \sum_{l=0}^{N-1} \mathcal{E}_{l} \ket{\phi_{l}}\bra{\phi_{l}},
\end{equation}
with $\{ \mathcal{E}_{0} \leq ... \leq \mathcal{E}_{N-1} \}$, and a given initial pure state $\ket{\psi_{0}}$, the maximal amount of energy that can be extracted 
(ergotropy) is given by \cite{allahverdyan2004maximal}
\begin{equation}
\label{eq:ergotropy_pure_states}
    \mathcal{W} = \mathcal{E}\left( \ket{\psi_{0}} \right) - \underset{\mathcal{U}}{\min}  \,\,\mathcal{E}\left( \mathcal{U} \ket{\psi_{0}} \right) \,,
\end{equation}
where $\mathcal{E} (|\psi_{0}\rangle) = \langle \psi_{0} |\mathcal{H} |\psi_{0} \rangle$.
Given an initial state, one has to find the unitary that minimize the energy of the system after
evolution. 

The ideal case is obtained when the initial state is an eigenstate 
with maximum energy and the state after the unitary is the ground state of the system
\begin{equation}
    \label{eq:max_ergotropy_pure_states}
    \mathcal{W}_{max} = \mathcal{E}\left( \ket{\phi_{max}} \right) - \mathcal{E}\left(  \ket{\phi_{min}} \right) = \mathcal{E}\left( \ket{\phi_{N-1}} \right) - \mathcal{E}\left(  \ket{\phi_{0}} \right) = \Delta,
\end{equation}
where $\Delta$ is the {\em bandwidth} of the system, i.e., the difference between 
the maximum and the minimum eigenvalue of the system's Hamiltonian. 

By writing the generic initial state in the Hamiltonian basis, one has
\begin{equation}
    \label{eq:initial_state}
    \ket{\psi_{0}} = \sum_{l=0}^{N-1} \alpha_{l} \ket{\phi_{l}}
\end{equation}
and the ergotropy reads
\begin{equation}
    \label{eq:ergotropy_pure_state_initial}
    \mathcal{W} = \mathcal{E}\left( \ket{\psi_{0}} \right) - \mathcal{E}\left(  \ket{\phi_{min}} \right) = \sum_{l=0}^{N-1} \abs{\alpha_{l}}^{2} \mathcal{E}_{l} - \mathcal{E}_{0}\,.
\end{equation}
The ideal unitary $\mathcal{U}$ for work extraction is thus the one taking any initial 
state $\ket{\psi_{0}}$ and projecting it into the ground state of the system 
$\ket{\phi_{min}} = \ket{\phi_{0}}$, i.e. 
\begin{equation}
    \label{eq:action_time_evolution}
    \mathcal{U} \ket{\psi_{0}} = \ket{\phi_{0}},
\end{equation}
Following this considerations, the operator $\mathcal{U}$ may  be written as 
\begin{equation}
   \label{eq:projector}
    \mathcal{U} = \ket{\phi_{0}}\bra{\psi_{0}} + \mathcal{B},
\end{equation}
where the matrix $\mathcal{B}$ is any matrix such that
\begin{equation}
   \label{eq:condition_B}
    \mathcal{B} \ket{\psi_{0}} = 0.
\end{equation}
The operator $\mathcal{U}$ is then highly non-unique, due to the degrees of freedom in the definition of $\mathcal{B}$. To find a form for $\mathcal{U}$, we can in principle consider any orthonormal basis which include $\ket{\psi_{0}}$, i.e. $\{ \ket{\psi_{l}}\}$, such that 
\begin{equation}
   \label{eq:orthonormality}
    \bra{\psi_{i}}\ket{\psi_{l}} = \delta_{il}\,,
\end{equation}
and then write the operator $\mathcal{U}$ as
\begin{equation}
   \label{eq:projector_generalized}
    \mathcal{U} = \sum_{l=0}^{N-1} \ket{\phi_{l}} \bra{\psi_{l}},
\end{equation}
where the non-uniqueness of $\mathcal{U}$ can be explicitly understood 
expanding the sum as
\begin{equation}
   \label{eq:projector_expanded}
    \mathcal{U} = \ket{\phi_{0}} \bra{\psi_{0}} + \sum_{l=1}^{N-1} \ket{\phi_{l}} \bra{\psi_{l}}\,.
\end{equation}
%

Let us now focus on building the optimal unitary as the time evolution under the 
action of an Hamiltonian given by the sum of the system's free Hamiltonian plus 
an external potential $\mathcal{V}(t)$.  We start from the Schrödinger equation
\begin{equation}
    \label{eq:Schrodinger}
    \partial_{t} \ket{\psi_{0}} = -i\mathcal{H}_{tot} \ket{\psi_{0}} = -i\left( \mathcal{H} + \mathcal{V}(t) \right) \ket{\psi_{0}}
\end{equation}
and impose that the time evolution operator of the system fulfills the condition
\begin{equation}
   \label{eq:time_evolution}
    \mathcal{T}\left( e^{-i\int_{0}^{t^{*}} \mathcal{H}_{tot} dt} \right) = \mathcal{U} = \sum_{l=0}^{N-1} \ket{\phi_{l}}\bra{\psi_{l}}.
\end{equation}
Since any realistic implementation of a time dependent potential is more prone to imperfections and errors, we look for a solution involving a time independent potential, with the projector condition reduced to
\begin{equation}
   \label{eq:time_independent_time_evolution}
    e^{-i \mathcal{H}_{tot} t^{*}} = \mathcal{U} = \sum_{l=0}^{N-1} \ket{\phi_{l}}\bra{\psi_{l}}.
\end{equation}
The total Hamiltonian should thus fulfill the condition
\begin{equation}
   \label{eq:condition_time_evolution}
    \mathcal{H}_{tot} = \frac{i}{t^{*}} \log \mathcal{U}  = \frac{i}{t^{*}} \log \left( \sum_{l=0}^{N-1} \ket{\phi_{l}}\bra{\psi_{l}} \right)\,,
\end{equation}
where
\begin{align}
   \label{eq:Taylor}
    \log \mathcal{U} =  \log \left[ \mathbf{1} + (\mathcal{U}- \mathbf{1}) \right]
    = \sum_{k=1}^{\infty} \frac{(-1)^{k+1}}{k!} (\mathcal{U}- \mathbf{1})^{k} =  - \sum_{k=1}^{\infty} \frac{(\mathbf{1}-\mathcal{U})^{k}}{k!}.
\end{align}
\subsection{Ergotropy for mixed states}
The concept of ergotropy can be generalized to the mixed state scenario, as
\begin{equation}
    \label{eq:ergotropy_non_pure}
    \mathcal{W} = \mathcal{E}\left( {\rho} \right) - \underset{\mathcal{U}}{\min} \,\,\mathcal{E}\left( \mathcal{U} \; {\rho} \; \mathcal{U}^{\dagger} \right) \,,
\end{equation}
where \textcolor{black}{$\mathcal{E}(\rho)=\Tr[\mathcal{H}\rho]$} and the optimization is again performed over the unitaries $\mathcal{U}$. Upon writing the density operator using its eigenbasis
\begin{equation}
    \label{eq:density_matrix_diag}
    {\rho} = \sum_{l=0}^{N-1} p_{l} \ket{\eta_{l}} \bra{\eta_{l}},
\end{equation}
one individuates the basis $\{ \ket{\eta_{l}} \}$ and the change of basis $\mathcal{D}$
\begin{equation}
    \label{eq:basis_change}
    \ket{\eta_{l}} = \sum_{k=0}^{N-1} \bra{\phi_{k}}\ket{\eta_{l}} \ket{\phi_{k}} = \sum_{k=0}^{N-1} \mathcal{D}_{l,k} \ket{\phi_{k}}.
\end{equation}
The optimal evolution of the system, i.e., that extracting the maximal amount of work from the system, is now given by
\begin{equation}
    \label{eq:ideal_transformation}
    \mathcal{U} = \mathcal{S}\mathcal{D}^{\dagger},
\end{equation}
where $\mathcal{D}$ is the matrix describing the above change of basis, and $\mathcal{S}$ is the unitary that rearrange the population probability of the density matrix $p_{s} = \bra{\phi_{s}} \rho \ket{\phi_{s}}$ 
in ascending order, thus leading to the output diagonal state
\begin{equation}
    \label{eq:final_state_density_matrix_diag}
    {\rho}_{\nearrow} = \sum_{l=0}^{N-1} p_{l,\nearrow} \ket{\phi_{l}}\bra{\phi_{l}}\,.
\end{equation}

\textcolor{black}{Following the same approach used for pure states, we now focus on constructing the optimal unitary by considering it as the time evolution generated by a Hamiltonian composed of the system’s free Hamiltonian plus an external potential $\mathcal{V}$ in the same form of Eq.\eqref{eq:time_independent_time_evolution}. We start from the Von-Neumann equation
\begin{equation}
    \label{eq:Von_Neumann}
    \partial_{t} \rho = -i\left[\mathcal{H}_{tot},\rho \right] = -i \left[ \mathcal{H} + \mathcal{V} ,\rho \right]\,,
\end{equation}
and impose that the time evolution operator of the system fulfills the condition of Eqs.\eqref{eq:ideal_transformation}-\eqref{eq:final_state_density_matrix_diag}, as
\begin{equation}
    \label{eq:time evolution non pure}
    {\rho}_{\nearrow} = e^{-i\mathcal{H}_{tot}t^{*}} \rho \, e^{i\mathcal{H}_{tot}t^{*}} \,.
\end{equation}
The total Hamiltonian should thus fulfill the condition 
\begin{equation}
   \label{eq:condition_time_evolution_non_pure}
    \mathcal{H}_{tot} = \frac{i}{t^{*}} \log \mathcal{U}  = \frac{i}{t^{*}} \log \left( \mathcal{S}\mathcal{D}^{\dagger} \right)\,.
\end{equation}}

The corresponding work extraction is then given by \cite{allahverdyan2004maximal}
\begin{align}
    \label{eq:ergotropy_non_pure_states}
    \mathcal{W} = \sum_{j,k=0}^{N-1} p_{j}\, \left( \vert \bra{\eta_{j}}\ket{\phi_{k}}\vert^{2} - \delta_{jk} \right)\,\mathcal{E}_{k}\,.
\end{align}
Of course,  Eq.\eqref{eq:final_state_density_matrix_diag} holds also for pure initial states, with only one element of the probability associated to $\ket{\phi_{0}}$ different from zero. The main difference between the 
pure and mixed case, is that for mixed states one cannot obtain a unitary evolution that project the  initial non-pure state onto the ground state of the system with a unitary matrix. This directly derive from the fact that not only the trace of ${\rho}$, but also the trace of ${\rho}^{2}$ must be conserved. Consequently, the transformation that saturates the ergotropy bound must be a rearrangement of the eigenvalues. 

In both the pure and mixed scenario, a state from which one cannot extract work is referred to as a \textit{passive} state. In other words, a state ${\zeta}$ is passive when $\hbox{Tr}[{\mathcal{H}{\zeta}}] \leq \hbox{Tr}[\mathcal{H}\mathcal{U}{\zeta}\mathcal{U}^{\dagger}]$ for all the possible unitaries $\mathcal{U}$. Following Eqs.\eqref{eq:ideal_transformation}-\eqref{eq:final_state_density_matrix_diag}, one has that 
${\zeta}$ is passive if and only if it is diagonal in the basis of the Hamiltonian $\mathcal{H}$, and its eigenvalues are non-increasing with the energy, i.e.,
\begin{equation}
    \label{eq:passive_state}
    {\zeta} = {\rho}_{\nearrow}.
\end{equation}
On the other hand, a generic state 
\begin{equation}
    \label{eq:active_state}
    {\aleph} = \sum_{e,f=0}^{N-1} {q}_{ef} \ket{\phi_{e}}\bra{\phi_{f}}.
\end{equation}
(with $q_{ef}=q_{fe}^{*}$), is referred to as thermodynamically active if
\begin{equation}
    \label{eq:energy_active_state}
    \langle \mathcal{E_{\aleph}} \rangle = \hbox{Tr}\left[\mathcal{H}{\aleph} \right]\geq  \hbox{Tr}\left[\mathcal{H}{\zeta}\right].
\end{equation}

\section{Non-Ideal Quantum Batteries}
\label{sec:FQB}
To develop realistic models of temporary quantum energy storage systems, it is also essential to describe the potential degradation mechanisms that worsen the battery's thermodynamic performance.
Let  us first consider a quantum system at thermal equilibrium. The state of the system
is time-invariant and it is described by the mixture
\begin{equation}
    \label{eq:thermal_rho}
    {\rho}_{th} = \frac{1}{\mathcal{Z}} \sum_{l=0}^{N-1} e^{-\beta\mathcal{E}_{l}} \ket{\phi_{l}}\bra{\phi_{l}}
\end{equation}
where $\beta$ is the inverse temperature $\beta=1/k_{B}T$ and $\mathcal{Z}=\Tr[e^{-\beta \mathcal{H}}]$ is the partition function. In this configuration the diagonal elements of the density matrix are in descending order, with the highest population associated to the lowest eigenvalue, and the state is then passive (see Eqs.\eqref{eq:final_state_density_matrix_diag}-
\eqref{eq:passive_state}). The battery thus needs to be recharged, before extracting any 
work. The ideal final state after the recharging procedure of a thermal state can be 
obtained from Eqs. \eqref{eq:final_state_density_matrix_diag} and \eqref{eq:thermal_rho} 
and reads
\begin{equation}
    \label{eq:inverse_thermal_rho}
    {\rho}_{th,inv} = \frac{1}{\mathcal{Z}} \sum_{l=0}^{N-1} e^{-\beta\mathcal{E}_{N-1-l}} \ket{\phi_{l}}\bra{\phi_{l}}.
\end{equation}
The subscript ``$inv$'' refers to the fact that the ideal unitary for recharging simply inverts the population of the energy levels from ascending to descending order. Due to this property, 
from now on, we will refer to a state in the form of Eq.\eqref{eq:inverse_thermal_rho} as the inverse thermal state. The ergotropy that can be extracted from such state can be obtained from Eq.\eqref{eq:ergotropy_non_pure_states} by substituting the exponentially decaying occupation probabilities of the density matrix of Eq.\eqref{eq:inverse_thermal_rho} and reads

\begin{equation}
    \label{eq:ergotropy_thermal_rho}
    \mathcal{W}_{th,inv} = \frac{1}{\mathcal{Z}} \sum_{l=0}^{N-1} (\mathcal{E}_{N-1-l} - \mathcal{E}_{l}) e^{-\beta\mathcal{E}_{l}} = \frac{1}{\mathcal{Z}} \sum_{l=0}^{N-1} \mathcal{E}_{l} \left(  e^{-\beta\mathcal{E}_{N-1-l}} - e^{-\beta\mathcal{E}_{l}} \right).
\end{equation}
In the limit of $\beta \rightarrow \infty$ (i.e. $T \rightarrow 0$) the density matrix ${\rho}_{th} \rightarrow \ket{\phi_{0}}\bra{\phi_{0}}$ and consequently ${\rho}_{th,inv} \rightarrow \ket{\phi_{N-1}}\bra{\phi_{N-1}}$, with the ergotropy in such limit that can be approximated as

\begin{equation}
    \label{eq:limit_ergotropy_thermal_rho}
    \lim_{T\rightarrow 0}\mathcal{W}_{th,inv} = \mathcal{W}_{max} = \Delta\,,
\end{equation}
which correctly coincides with the result obtained for ideal work extraction from 
a pure state.

The battery may be in a mixed state also because subject to non unitary evolution due 
to the presence of any type of intrinsic or extrinsic decoherence and dissipation \cite{manzano2020short,bressanini2022decoherence,Farina2019,Morrone2023}.  The evolution in a noisy 
environment is described by a master equation
\begin{equation}
    \label{eq:noisy_evolution}
    \partial_{t} {\rho}(t) = -i\left[ \mathcal{H},{\rho}(t) \right] + \sum_{k} \gamma_{k} \mathcal{D}[O_{k}]{\rho}(t)\,.
\end{equation}
where $\mathcal{D}[O_k]$ denotes the $k$-th Lindblad operator of the system, 
with action defined by $\mathcal{D}[O_k]\rho(t) =O_k \rho(t) O_k^{\dagger} - \frac{1}{2} 
\{ O_k^{\dagger}O_k,\rho(t) \}$,  where $\{A,B\}$ denotes the anticommutator between the operators $A$ and $B$, the $D_k$ are jump operators, describing the dissipative part 
of the dynamics, and $\gamma_{k} \geq 0$ is the noise rate of the $k$-th decay channel.

Pure dephasing, i.e., the loss of coherence in the energy basis,  \cite{milburn1991intrinsic,schlosshauer2019quantum}  corresponds to the
following master equation 
\begin{equation}
    \label{eq:decoherence_energy_basis}
    \partial_{t} {\rho}(t) = -i\left[ \mathcal{H},{\rho}(t) \right] - \frac{\gamma}{2} \left[\mathcal{H}, \left[ \mathcal{H},{\rho}(t) \right] \right],
\end{equation}
which describes the exponential suppression of the off-diagonal matrix elements at
rate $\gamma \geq 0$.

Another kind of noisy quantum evolution corresponds to decoherence in position basis,
described by the Haken–Strobl master equation \cite{haken1973exactly,hoyer2010limits}
\begin{equation}
    \label{eq:haken_strobl}
    \partial_{t} {\rho}(t) = -i\left[ \mathcal{H},{\rho}(t) \right] + \gamma \sum_{k} \mathcal{D}\left[P_{k}\right] {\rho}(t),
\end{equation}
where $\gamma \geq 0$ again denotes the decoherence rate and $P_{k} = \ket{k}\bra{k}$ are  projection operators over discrete position states.  The equation can be interpreted as the 
interaction of the system with a fluctuating environment \cite{rebentrost2009environment,sarovar2011environmental} inducing localization. 

Finally, as a last decoherence model, we consider the so-called quantum stochastic walk (QSW) model \cite{kempe2003quantum,whitfield2010quantum,caruso2014universally}
\begin{equation}
    \label{eq:quantum_stochastic_walk}
    \partial_{t} {\rho}(t) = -i(1-p)\left[ \mathcal{H},{\rho}(t) \right] + p \sum_{kj} \mathcal{D}\left[P_{kj}\right] {\rho}(t),
\end{equation}
in which $p \in \left[0,1\right]$ quantifies the interplay between the standard Hamiltonian evolution $(p = 0)$ and the irreversible dynamics $(p = 1)$ and with $P_{kj}= \bra{k}\mathcal{H}\ket{j} \ket{k}\bra{j}$ that represent the transition operators.

\section{Ideal Potential for Work Extraction}
\label{sec:IpfWE}
Let us now devote attention to the design of potentials possibly maximizing the 
ergotropy for a large class of quantum systems under different noise conditions, 
independently of the number of energy levels. We consider two distinct efficient 
protocols able to saturate the ergotropy limit, both based on a time dependent 
potential $\mathcal{V}(t)$ in the form
\begin{equation}
    \label{eq:ideal_potential}
    \mathcal{H}_{tot} = \mathcal{H} + \mathcal{V}(t) = \mathcal{H} + \left( \Theta(t)-\Theta(t-t^{*}) \right)\mathcal{V},
\end{equation}
The first ideal charging-discharging protocol for a general quantum battery may 
be achieved by a potential $\mathcal{V}(t)$ that leads to a piecewise Hamiltonian 
of the form
\begin{equation}
    \label{eq:ideal_potential_time_intervals}
    \mathcal{H}_{tot}= \begin{cases}
        \sum_{l=0}^{N-1} \mathcal{E}_{l} \ket{\phi_{l}}\bra{\phi_{l}}\,, 
        \qquad \qquad \qquad \qquad\qquad t<0\,,   \\
         \xi\, \sum_{l=0}^{N-1} \frac{1}{2}\sqrt{l(N-l)}  \ket{\phi_{l}}\bra{\phi_{l+1}} + h.c. , \; \; \; \; 0<t<t^{*}\,,\\
         \sum_{l=0}^{N-1} \mathcal{E}_{l} \ket{\phi_{l}}\bra{\phi_{l}}\,, 
        \qquad \qquad \qquad \qquad\qquad t>t^{*}\,.
    \end{cases}
\end{equation}
The corresponding time evolution operator $\mathcal{U}$, evaluated at $t^{*}=\pi/\xi$, with $\xi$ being the coupling in the time interval  $0<t<t^{*}$, reads as
\cite{christandl2004perfect,christandl2005perfect,yung2005perfect,cavazzoni2025perfect}
\begin{equation}
    \label{eq:ideal_projector_from_time_evolution}
    \mathcal{U}(\pi/\xi)=e^{-i\mathcal{H}\pi/\xi} = \sum_{l=0}^{N-1} \ket{\phi_{l}} \bra{\phi_{N-1-l}},
\end{equation}

A second optimal protocol is given by considering a potential $\mathcal{V}(t)$ leading to a piecewise Hamiltonian of the form 
\begin{equation}
    \label{eq:ideal_potential_time_intervals_2}
    \mathcal{H}_{tot}= 
    \begin{cases}
      \sum_{l=0}^{N-1} \mathcal{E}_l \ket{\phi_l} \bra{\phi_l} \,, \qquad  \qquad t<0  \\
\chi\,\sum_{l=0}^{N-1}               \ket{\phi_l} \bra{\phi_{N-1-l}} \,, \;\; \;\; \; 0<t<t^{*} \\
      \sum_{l=0}^{N-1} \mathcal{E}_{l} \ket{\phi_l}\bra{\phi_l}\,, \qquad  \qquad  t>t^{*}.
      
    \end{cases}
\end{equation}
this leads to a time evolution operator $\mathcal{U}^{'}$ 
\begin{equation}
    \label{eq:ideal_projector_from_time_evolution_2}
    \mathcal{U}^{'}(t)=e^{-i\mathcal{H}_{tot}t} = \mathbb{I}\cos{(\chi t)}-i\sin{(\chi t)}\sum_{l=0}^{N-1} \ket{\phi_{l}} \bra{\phi_{N-1-l}},
\end{equation}
which at time $t^{*}=\pi/2\chi$, with $\chi$ being the coupling in the time interval  $0<t<t^{*}$, up to a global phase reduces to 
\begin{equation}
    \label{eq:ideal_projector_from_time_evolution_2_pi}
    \mathcal{U}^{'}\left(\pi/2\chi\right)= \sum_{l=0}^{N-1} \ket{\phi_{l}} \bra{\phi_{N-1-l}},
\end{equation}
which acts exactly as $\mathcal{U}(\pi/\xi)$. 

Both these potentials are useful since the ideal initial state is 
$\rho = \ket{\phi_{N-1}}\bra{\phi_{N-1}}$ and both the operators 
$\mathcal{U}$ and $\mathcal{U}^{'}$ drive the system towards 
its ground state, being the ideal transformation for work extraction. 
In the same way, when the system is defined by a density matrix in the form $\rho = \ket{\phi_{0}}\bra{\phi_{0}}$ the operator $\mathcal{U}$ defines the recharge process, as it drives the system towards the eigenstate with maximum energy. Accordingly, to recharge 
a system initially at thermal equilibrium, see Eq. (\ref{eq:thermal_rho}), the operator $\mathcal{U}^{-1}=\mathcal{U}$ ($\mathcal{U}^{'}$ or $\mathcal{U}^{',-1}$) drives the system towards the state with maximum energy that is accessible through a unitary transformation. 
In the very same way $\mathcal{U}^{-1}=\mathcal{U}$ ($\mathcal{U}^{'}$ or 
$\mathcal{U}^{',-1}$) can be adopted to extract the energy from the system to 
bring back the system towards its thermal passive state. 

As a matter of fact, the unitaries $\mathcal{U}$ and $\mathcal{U}^{'}$ are suitable to
extract maximally energy and recharge a battery in a generic mixed state. 
Indeed, the ideal unitary transformation for charging a quantum system defined by a density matrix with population in descending (ascending) order in energy basis are $\mathcal{U}$ and $\mathcal{U}^{'}$. On the other hand, if the state does not have the population in ascending or descending order in energy basis, the ideal operator $\Bar{\mathcal{U}}$, associated to such state is just a rearrangement of the lines of $\mathcal{U}$ or $\mathcal{U}^{'}$.

\section{Continuous-time quantum walks}
\label{sec:QWT}
Let us now address the definition of the Hamiltonian of the system. Our model is based on continuous time quantum walk (CTQW), which describes a quantum system (the walker) 
confined on a discrete set of spatial locations, defining the set of vertices of a graph. 
In CTQWs, the generator of the dynamics, i.e. the Hamiltonian $\mathcal{H}$, is intrinsically defined by the topology of the discrete space and by its connection. Graph theory provides the mathematical framework for the definition of $\mathcal{H}$ through the arrangement of the discrete sites (vertices) and their connections (edges). Specifically, a graph is an ordered pair $G = \left( V, E \right)$ where $V$ is the set of vertices and $E$ the set of edges. The adjacency matrix $\mathcal{A}$ is defined by the connection of the vertices and its elements $\mathcal{A}_{jk}$ are $\mathcal{J}$ if and only if the two vertices $j$, $k$ are connected, 0 otherwise, as 
\begin{equation}
\mathcal{A}_{jk} = \begin{cases}
\mathcal{J} & \text{if  $j \neq k$ and $(j, k) \in E$,}\\
0 & \text{otherwise,} 
\end{cases}
\end{equation}
and the value of $\abs{\mathcal{J}}$ defines the intensity of the coupling among the discrete sites. The Hamiltonian of such system is then obtained through the adjacency matrix as $\mathcal{H}=-\mathcal{A}$. The vertices of the graphs represent the possible positions of the walker, whereas the edges correspond to the possible transitions among the vertices. The states $\left\{\ket{j}\right\}_{j=0,...,N-1}$, where $j \in V$ is a vertex of $G$, form an orthonormal basis and span the $N$-dimensional Hilbert space of the quantum walker. In principle, a CTQW is generated by any Hamiltonian $\mathcal{H}$ (or, generally, any Hermitian operator) that respects the topology of the graph, promoting the adjacency matrix to a valid generator for quantum dynamics. Due to the intrinsic quantum nature of the model, the adjacency matrix can also be defined through complex entries \cite{lu2016chiral} as
\begin{equation}
\mathcal{C}_{jk} = \begin{cases}
\mathcal{J}e^{-i\phi_{jk}} & \text{if  $j \neq k$ and $(j, k) \in E$,}\\
0 & \text{otherwise,} 
\end{cases}
\end{equation}
with $\mathcal{C}_{jk} = \mathcal{C}_{kj}^{*}$, intrinsically providing additional quantum characteristics to the dynamics of the system. In such a way the Hamiltonian of the system is generalized to a matrix with complex entries as $\mathcal{H}=\mathcal{C}$.

By modeling quantum batteries with CTQW, one can investigate how structural properties, such as connectivity, affect the thermodynamical features of the system and enable. In addition, the presence of complex phases in the off-diagonal elements of the Hamiltonian, breaks the symmetry under the time-reversal transformation $t \rightarrow -t$, 
i.e., introduces chirality in the system. 
\subsection{Dimensionality reduction method}
In quantum mechanics, dimensionality reduction techniques are often used to simplify 
the description of high-dimensional quantum states and processes while preserving 
essential features \cite{caruso2009highly,novo2015systematic}. For CTQWs, dimensionality reduction methods are used 
to simplify the calculation of the generic transition amplitude of the initial state $|\psi_0\rangle $
to a given location $a$ at a given time $t$, i.e., $\braket{a}{\psi(t)}$.
Indeed, upon exploiting both the symmetries of the state and of the graph, it is possible to map the original problem to an equivalent one, within a reduced Hilbert space dimension. Considering the Taylor expansion of the time-evolution operator, the transition amplitude 
$\bra{a}e^{-i \mathcal{H}t}\ket{\psi_{0}}$ can be written as
\begin{align}
\label{eq:Taylor_Krylov}
\bra{a}e^{-i \mathcal{H}t}\ket{\psi_{0}}&=\sum_{k=0}^\infty \frac{(-it)^k}{k!}\bra{a}\mathcal{H}^k\ket{\psi_{0}}\nonumber\\
&=\bra{a}e^{-i\mathcal{H}_\textup{red} t}\ket{{\psi_{0}}_\textup{red}}\,,
\end{align}
where  $ \mathcal{PHP}=\mathcal{H}_\textup{red} $ is a reduced Hamiltonian, and $\ket{{\psi_0}_\textup{red}}= \mathcal{P}\ket{\psi_0}$ a reduced initial state, $\mathcal{P}$ being the projector onto the so-called Krylov subspace, which itself is defined as
\begin{equation}
\mathcal{I}(\mathcal{H},\ket{\psi_{0}}) = \operatorname{span}(\lbrace \mathcal{H}^k \ket{\psi_{0}} \mid k \in \mathbb{N}_0\rbrace)\,.
\label{eq:Krylov_subspace}
\end{equation}

An orthonormal basis, $\{\ket{e_0},\ldots,\ket{e_{m-1}}\}$, for the Krylov subspace $\mathcal{I}(\mathcal{H},\ket{w})$ can be iteratively constructed starting from the state $\ket{e_0}=\ket{\psi_{0}}$, by applying $\mathcal{H}$ and orthonormalizing the result with 
Gram-Schmidt procedure. The procedure ends when the state $\mathcal{H}\ket{e_{m-1}}$ is a linear combination of the previous states $\ket{e_0},\ldots,\ket{e_{m-2}}$. The resulting reduced Hamiltonian written in such a basis has a tridiagonal form. At each iteration, the state $\mathcal{H}\ket{e_k}$ is a linear combination of $\ket{e_{k-1}}$, $\ket{e_{k}}$, and the new basis state $\ket{e_{k+1}}$ to be defined. The original problem is then mapped onto an equivalent one of lower dimension, governed by a tight-binding Hamiltonian on a line (path graph) with $m$ sites, with $m$ generally lower than $N$. Then, on such basis the elements of the Hamiltonian $\mathcal{H}_{red}$ of the system are written as

\begin{equation}
    \label{eq:H_red}
    \mathcal{H}_{red \, j,k} = \bra{e_{j}}\mathcal{H}\ket{e_{k}}\,.
\end{equation}

\section{A larger view: from  Q-Cells to  battery engineering}
\label{sec:ALV}
In our model, a quantum battery is composed by $\mathcal{M}$ quantum cells (Q-Cells), each one 
corresponding to a quantum walker on a graph, which represent the fundamental 
building blocks of the battery. The overall battery is thus described by 
the factorized density operator $\rho_{b}$
\begin{equation}
    \label{eq:multiple_cells_battery}
    \rho_{b} = \bigotimes^{\mathcal{M}} \rho_{q-c}\,,
\end{equation}
where $\rho_{q-c}$ denotes the state of a single cell.
The global Hamiltonian of the battery is given by
\begin{equation}
    \label{eq:global_Hamiltonian}
    \mathcal{H}^{(\mathcal{M})} = \sum_{l=1}^{\mathcal{M}} \mathcal{H}_{l},
\end{equation}
where $\mathcal{H}_{l}$ is defined as
\begin{equation}
    \label{eq:single_Hamiltonian}
    \mathcal{H}_{l} := \mathbb{I}_{1} \otimes ... \mathbb{I}_{l-1} \otimes \mathcal{H} \otimes \mathbb{I}_{l+1} \otimes ... \mathbb{I}_{\mathcal{M}}.
\end{equation}

\begin{figure}[!ht]
  \centering
  \includegraphics[width=0.4\columnwidth]{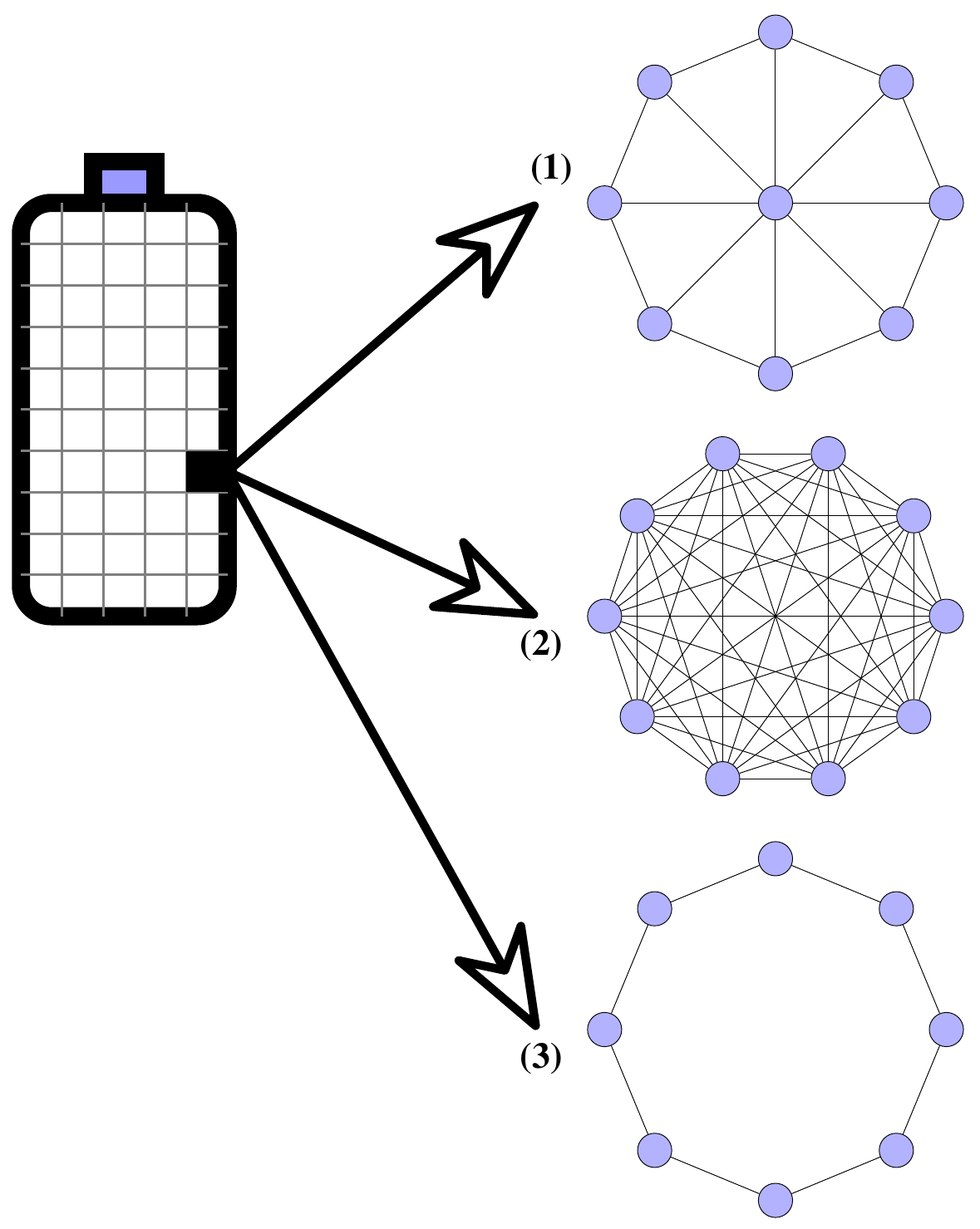}
\caption{Schematic representation of a quantum battery. On the left the graphical representation of the whole battery, which consist in an arrangement of Q-Cells. On the right the focus on the structure of the topology of the Q-Cells analyzed: (1) Wheel Q-Cell, (2) Complete Q-Cell and (3) Ring Q-Cell.}
\label{fig:unit_cell}
\end{figure}

The schematic structure of the whole battery is reported in Fig.\ref{fig:unit_cell}. The overall battery is a collection of Q-Cells that correspond to a general discrete structure defined by a quantum walk model. Specifically, we will focus on three different type of networks: (1) the ring Q-Cell, in which the elements are arranged in ring and there is only nearest neighbors interaction, (2) the complete Q-Cell in which each element of the structure interacts with all the others and (3) the wheel Q-Cell in which each element of a ring structure interacts with a central element.

The operation aimed to extract energy from the battery may be basically divided into two main groups: local Q-Cell operations and global battery operations. The first group of operations involves separately every single Q-Cell, defined by a potential $\mathcal{V}_{l}(t)$, as
\begin{equation}
    \label{eq:single_potential}
    \mathcal{V}_{l}(t) := \mathbb{I}_{1} \otimes ... \mathbb{I}_{l-1} \otimes \mathcal{V}(t) \otimes \mathbb{I}_{l+1} \otimes ... \mathbb{I}_{\mathcal{M}}.
\end{equation}
In such a way, the global ergotropy of the whole battery may be simply obtained as

\begin{equation}
    \label{eq:global_ergotropy}
    \mathcal{W}_{b} = \sum_{l=1}^{\mathcal{M}} \mathcal{W}_{q-c} = \mathcal{M} \cdot \mathcal{W}_{q-c}
\end{equation}
The final state of the whole battery, after the transformation defined in Sec.\ref{sec:QWEiaN}, is a tensor product of $\mathcal{M}$ passive states, and do not correspond in general to a globally passive state. The total time required to perform a procedure to extract energy from the battery is the same as the time for a single Q-Cell $t^{*}_{b}=t^{*}_{q-c}$, as the operations can be performed in parallel.

The second set of possible operations, involves global operations that acts on the whole battery, possibly entangling cells, rather than on each cell separately \cite{campaioli2017enhancing}. In such condition a quantum advantage may arise \cite{gyhm2022quantum} in terms of charge and discharge time, i.e., it concerns the power of the battery. In other words, the ergotropy is equal to that in Eq.\eqref{eq:global_ergotropy}, but the time required to perform the optimal global unitary may be different, leading to different power of the battery. In the following, we will focus on achieving the ergotropy limit and on the study of the thermodynamical properties of the single cells under chirality and noise effects.  We therefore focus attention on local Q-Cell operations.

\section{The charge-discharge cycle}
\label{sec:TCDC}
\begin{figure}[!ht]
  \centering
  \includegraphics[width=0.7\columnwidth]{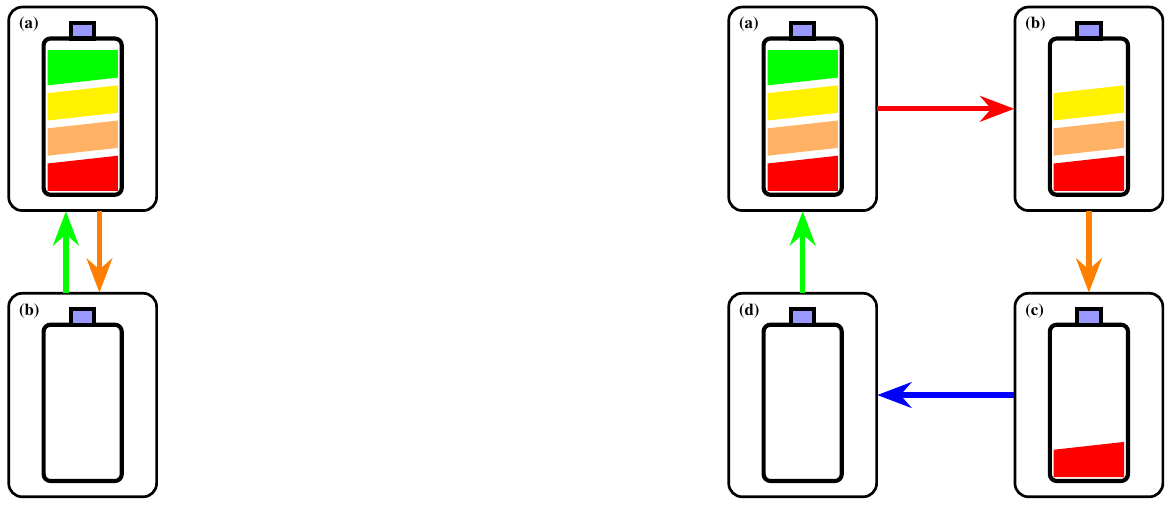}
    \caption{Schematic representation of the charge-discharge cycle of a quantum battery, where each panel corresponds to a different battery state. 
Left panel (a): Fully charged battery (described by the pure eigenstate $\ket{\psi}=\ket{\phi_{N-1}}$); Left (b): Ground state of the battery (described by the pure eigenstate $\ket{\psi}=\ket{\phi_{0}}$). Right panel (a): Fully charged battery (described by the pure eigenstate ${\rho}=\ket{\phi_{N-1}}\bra{\phi_{N-1}}$). Right (b:) State of the battery after a non-ideal evolution (corresponding to the density matrix ${\rho}(\tau)$). Right (c): Passive state of the battery (corresponding to the state ${\zeta}$ introduced in Eq.\eqref{eq:passive_state}). Right (d): Ground state of the battery (described by the pure eigenstate ${\rho}=\ket{\phi_{0}}\bra{\phi_{0}}$).}
    \label{fig:cycle_battery}
\end{figure}

In this section we illustrate in details the charging and discharging of the 
battery \cite{qi2025quantum}. For pure states the first step, see left panel (a) of Fig.\ref{fig:cycle_battery}, is to initialize the battery in the eigenstate associated to the highest eigenvalue, $\ket{\psi}=\ket{\phi_{N-1}}$. The discharge operation consist in the projection of the initial state onto the ground state of the battery, passing from 
$\ket{\psi}=\ket{\phi_{N-1}}$ to $\ket{\psi}=\ket{\phi_{0}}$, see left panel (b) of 
Fig.\ref{fig:cycle_battery}. The charging process consist in the opposite procedure, 
inducing the transition from the ground state of the battery to its highest energy 
state.

The first step for mixed states, see the right panel (a) of Fig. \ref{fig:cycle_battery}, coincides with the first step for pure states and consist in initializing the battery in the eigenstate associated to the highest eigenvalue, $ {\rho}= \ket{\phi_{N-1}}\bra{\phi_{N-1}}$. Then, the energy remain stored for a time $\tau$, before the application of the discharge operation. Due to the interaction with the environment the overall energy available in the battery is reduced, see right panel (b). At this point the battery is described by a mixed state through the density matrix $\rho(\tau)$. To exploit the energy that is still present in the battery it is necessary to apply the $\mathcal{U}$ transformation introduced in Eq.\eqref{eq:ideal_transformation} that drives the battery to a passive state, see panel (c). 
At this point, it is possible to apply a projective measurement to steer the battery towards its ground state, see panel (d). To recharge the battery, it is necessary to implement the transition from its ground state to its maximally energetic eigenstate 
(the pure state scenario may, in turn, be seen as a limit of the mixed state one, 
where the discharge of the battery happens immediately after the preparation in its maximally energetic state). Then, the cycle can begin again, repeating in order the same steps.

Concerning the specific steps of the cycle, few remarks have to be done. The non-ideal evolution that discharge the battery (transition from panel (a) to panel (b) in Subfig.\ref{fig:cycle_battery}.2) is non unitary, as in general it changes the eigenvalue of the density matrix. In the same way, the process adopted to drive the battery to its ground is again non unitary, since it is a transition from a mixed  to a pure state (from panel (c) to panel (d) in Subfig.\ref{fig:cycle_battery}.2).  In particular,  
the projective measurement on the ground state is necessary to fully exploit the quantum battery. Otherwise, using only unitaries (especially in the mixed case), the state of the battery is restricted only to panels (b) and (c) of Fig.\ref{fig:cycle_battery}, and the ergotropy of the system is lower. 
\section{The Wheel Q-Cell}
\label{sec:SB}
As a first model, we discuss the performance of a battery composed of wheel-structured Q-Cell, corresponding to the interaction of a single discrete element $(\ket{x_{0}})$ to an external 
ring of elements, forming the 1-skeleton of an (N–1)-gonal pyramid. Such structure also known 
as wheel graph has an associated (adjacency) Hamiltonian of the form
\begin{equation}
    \label{eq:Wheel Hamiltonian}
    \mathcal{H}^{w} = -\mathcal{J}\left( \sum_{l=1}^{N-1} \ket{x_0}\bra{x_l} + \sum_{l=1}^{N-1} \ket{x_l}\bra{x_{l+1}} \right) + h.c. ,
\end{equation}
where $\ket{x_{0}}$ is the central vertex, $\ket{x_{l}}$ is an element of the external ring, 
and $\ket{x_{N}}=\ket{x_{1}}$. The eigenvalues are given by
\begin{align}
    \label{eq:eigenvalues_wheel}
    &\mathcal{E}^{w}_{0,N-1}=-(1\pm \sqrt{N})\mathcal{J}, \nonumber \\
    &\mathcal{E}^{w}_l = -2\mathcal{J} \cos{\left( \frac{2\pi l}{N-1} \right)}, \,\,\,  l=1,2,...,N-2
\end{align}

\begin{figure}[!ht]
  \centering
  \includegraphics[width=0.3\columnwidth]{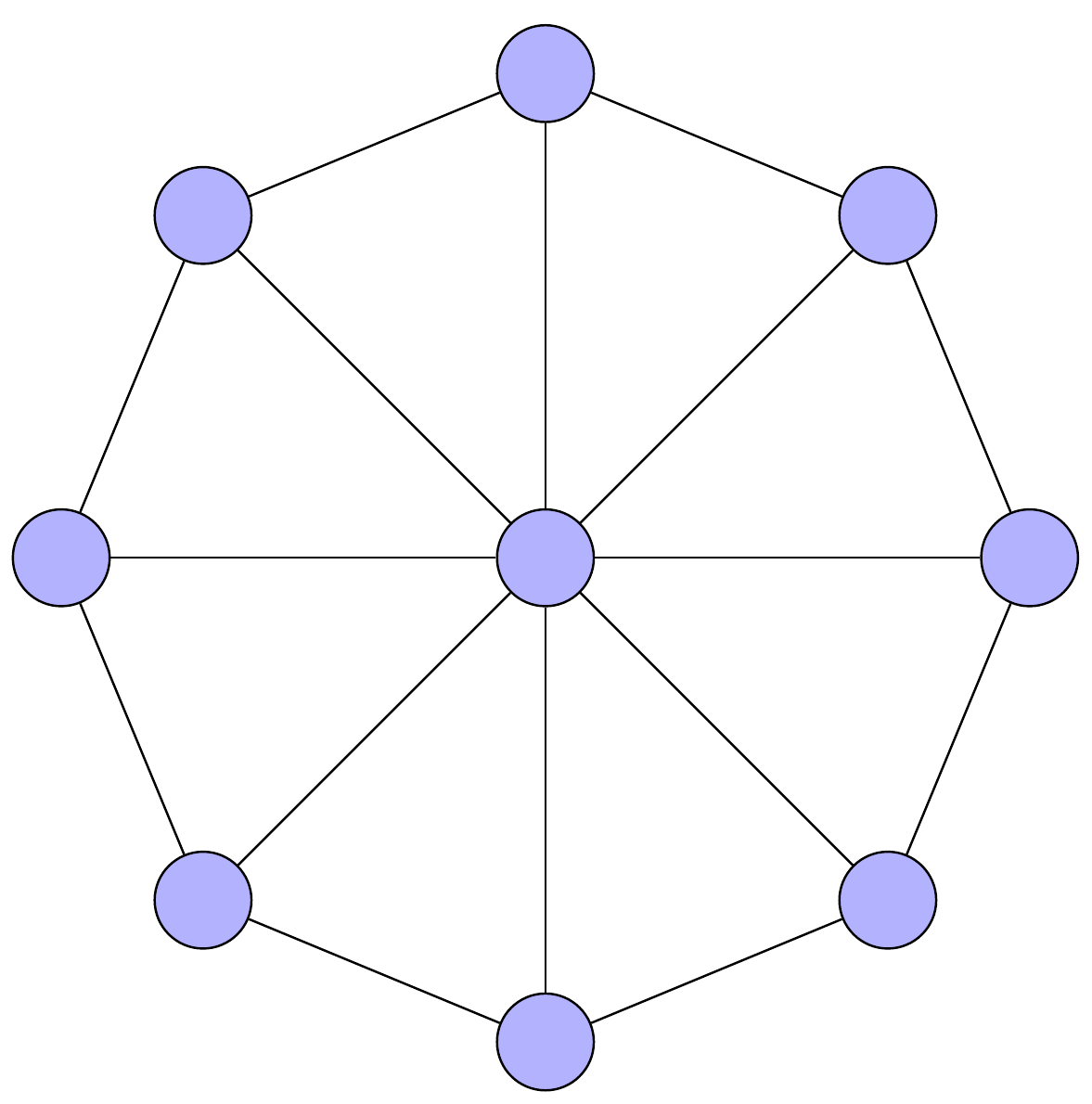}
\caption{Graphical representation of the Wheel Q-Cell composed of $N=9$ discrete elements for a skeleton quantum battery.}
    \label{fig:Wheel Battery}
\end{figure}

Following Eq. \eqref{eq:max_ergotropy_pure_states}, the maximal work that can be extracted from a wheel structure is
\begin{equation}
    \mathcal{W}_{max}^{w} = 2\sqrt{N}\mathcal{J}
\end{equation}
and grows as the square root of the size of the structure itself. An operator able to 
perform the corresponding transformation may be written as
\begin{equation}
    \label{eq:U_wheel_cell}
    \mathcal{U} = \sum_{l=0}^{N-1} \ket{\phi_{l}}\bra{\phi_{N-1-l}}\,.
\end{equation}

Additionally, as a paradigmatic example of non ideal initial state, we consider a state $\ket{\psi_{0}}=\ket{x_{0}}$ localized in the central vertex of the structure. The ergotropy for such state is
\begin{equation}
    \mathcal{W}_{loc}^{w} = (\sqrt{N}+1)\mathcal{J},
\end{equation}
calculated as the difference between the energy of such state $\langle\mathcal{E}_{x_{0}}\rangle=0$ and the ground state of the structure (i.e. $\mathcal{E}^{w}_{0}=-(1+\sqrt{N})$). Then, also the ergotropy for a localized state depends on the dimensionality of the system. As defined in Sec.\ref{sec:QWEiaN}, for a localized state in the form $\ket{\psi_{0}}=\ket{x_{0}}$ a possible ideal transformation for the maximal work extraction is
\begin{equation}
    \label{eq:U_matrix_wheel_cell_loc}
    \mathcal{U} = \ket{\phi_{0}}\bra{x_{0}} + \sum_{l=1}^{N-1} \ket{\phi_{l}}\bra{x_{l}}.
\end{equation}
The localized state can be adopted as a convenient state to initialize the battery. After the first discharge cycle, the battery will end up in the ground state, and then, the unitary transformation illustrated in Eq.\eqref{eq:U_wheel_cell} will fully recharge the battery. Then, the charge discharge cycle follows the exact same steps depicted in the highest lowest eigenstate transition.

Since the ergotropy increases with the dimensionality of the cell, let us analyze 
Q-Cell in the limit of a large size, with the application of an approximated 
transition from an initial localized state to ground state which works specifically 
for a large number of elements $N$. The dimensionality reduction method for a 
localized state $\ket{\psi_{0}}=\ket{x_{0}}$ in the wheel Q-Cell, following Eqs.\eqref{eq:Taylor_Krylov}-\eqref{eq:Krylov_subspace}, lead to a Krylov space in the form

\begin{equation}
\mathcal{I}(\mathcal{H}^{w},\ket{\psi_{0}}) = \{ \ket{\psi_{0}},\ket{s} \}
\label{eq:Krylov_subspace_w_q-c}
\end{equation}
where $\ket{s}$ is the equal superposition of all the discrete elements in the Q-Cell except $\ket{\psi_{0}}=\ket{x_{0}}$, as

\begin{equation}
\label{eq:ket_s_w_q-c}
\ket{s} = \frac{1}{\sqrt{N-1}} \sum_{l=1}^{N-1} \ket{x_l}.
\end{equation}
In such basis the Hamiltonian of the Q-Cell presented in Eq.\eqref{eq:Wheel Hamiltonian} is rewritten as

\begin{equation}
    \label{eq:Hamiltonian_wheel_red}
    \mathcal{H}^{w}_{red} = -\mathcal{J}\left(
    \begin{array}{cc}
        0 & \sqrt{N-1} \\
        \sqrt{N-1} & 2  
    \end{array}\right),
\end{equation}
and with the application of a localized potential $\mathcal{V}=-4\mathcal{J} \ket{\psi_{0}}\bra{\psi_{0}}$ in the system is induced the transition from the localized state to the ground state, which is the superposition of all the discrete position of the structure, as

\begin{equation}
    \label{eq:large cell limit method wheel}
    e^{-i(\mathcal{H}^{w}_{red}+\mathcal{V})t^{*}}\ket{\psi_{0}} \approx  \frac{1}{\sqrt{N}} \sum_{l=0}^{N-1} \ket{x_{l}}=\ket{\phi_{0}}.
\end{equation}

Let us now consider the charge discharge cycle for a wheel Q-Cell in thermal equilibrium. In such condition the state of the battery is defined through a density matrix in the form of Eq.\eqref{eq:thermal_rho}. Accordingly, when the system is in an inverse thermal state (see Eqs.\eqref{eq:inverse_thermal_rho}-\eqref{eq:ergotropy_thermal_rho}) the ergotropy reads

\begin{align}
    \label{eq:ergotropy_thermal_wheel}
        \mathcal{W}_{th,inv}^{w}=& \frac{ \textcolor{black}{ -2\mathcal{J} \sum_{l=1}^{N-2} \left( \cos{\left(\frac{2\pi}{N-1} f(l,N,\mathcal{K})\right)} - \cos{\left(\frac{2\pi (N-1-l)}{N-1} \right)} \right) e^{2\mathcal{J}\beta\cos{\left(\frac{2\pi (N-1-l)}{N-1} \right)}} } }{ 2e^{\beta \mathcal{J}} \cosh{\left( \beta \mathcal{J} \sqrt{N} \right)} + \sum_{l=1}^{N-2} e^{2\mathcal{J}\beta\cos{\left(\frac{2\pi l}{N-1} \right)}} }  +\nonumber \\
        &\frac{\left( 4\sqrt{N}\mathcal{J} e^{\beta \mathcal{J}} \sinh{\left( \beta \sqrt{N} \mathcal{J} \right)} \right)}{ 2e^{\beta \mathcal{J}} \cosh{\left( \beta \mathcal{J} \sqrt{N} \right)} + \sum_{l=1}^{N-2} e^{2\mathcal{J}\beta\cos{\left(\frac{2\pi l}{N-1} \right)}}}
\end{align}
where here we have introduced \textcolor{black}{$f(l,N,\mathcal{K})$ which reads
\begin{equation}
    \label{eq:f}
    f(l,N,\mathcal{K}) = \begin{cases}
        \mathcal{K}-1+l, \,\,\, if \,\,\, \mathcal{K}-1+l < N-1 \nonumber \\
        \mathcal{K}+1+l-N, \,\,\, if \,\,\, \mathcal{K}-1+l \geq N-1
    \end{cases}
\end{equation}
and $\mathcal{K}=N/2$ or $\mathcal{K}=(N-1)/2$ depending on the odd or even number of elements in the external ring of the cell.}

\section{Circulant Quantum Batteries}
\label{sec:CB}

Let us now focus on the case of batteries with Q-Cells given by circulant 
graphs (e.g., complete and ring battery Q-Cells) and on 
paradigmatic initial states. A general circulant Hamiltonian 
$\mathcal{H}_{\Gamma}$ in the basis $\{\ket{x_{m}}\}$, is represented by a matrix in the form

\begin{equation} 
    \label{eq:circulant_matrix}
    \mathcal{H}_{\Gamma} = 
    \left(
    \begin{array}{ccccc}
        q_0 & q_1 & q_2 & ... & q_{n-1}\\
        q_{n-1} & q_0 & q_1 & ... & q_{n-2} \\
        \vdots &\ddots & \ddots & \ddots & \vdots\\
        \vdots &\ddots & \ddots & \ddots & \vdots\\
        q_1 & q_2 & q_3 & ... & q_0
    \end{array}\right).
\end{equation}
In all the quantum systems represented by such matrices the eigenvectors coincides, and are written as

\begin{equation}
    \label{eq:eigenvectors_circulant}
    \ket{\phi_{l}} = \frac{1}{\sqrt{N}} \sum_{m=0}^{N-1} w^{m l} \ket{x_{m}}
\end{equation}
with $w=exp(\frac{i2\pi}{N})$ and $\{\ket{x_{m}}\}$ represent the position basis.
Associated to these eigenvectors, the corresponding eigenvalues are
\begin{equation}
\label{eq:eigenvalues_circulant}
    \mathcal{E}_l= \sum_{m=0}^{N-1} q_m \omega^{m l}
\end{equation}
 For any thermodynamical purpose the knowledge of the ground and maximally energetic state of the system is fundamental, and for circulant matrices such states are written as

\begin{equation}
    \label{eq:ground_state_circulant}
    \ket{\phi_{N-1}} = \frac{1}{\sqrt{N}} \sum_{m=0}^{N-1} w^{m(N-1)} \ket{x_{m}} \, ; \, \ket{\phi_{0}} = \frac{1}{\sqrt{N}} \sum_{m=0}^{N-1} \ket{x_{m}}
\end{equation}
For practical purposes, related to experimental implementations, we will investigate a localized initial state $\ket{\psi_{0}}=\ket{x_{m}}$, which in terms of the eigenvectors of the circulant matrices reads

\begin{equation}
\label{eq:loc_state}
    \ket{x_{m}} = \frac{1}{\sqrt{N}}\sum_{l=0}^{N-1} w^{-m l} \ket{\phi_{l}}\,.
\end{equation}
Due to the structure of Eq.\eqref{eq:loc_state} any localized state differs only by a phase factor in its eigenstate decomposition, every localized state in a circulant battery has an energy

\begin{equation}
\label{eq:en_loc_state}
    \langle \mathcal{E}_{x_{m}} \rangle= \frac{1}{N} \sum_{l=0}^{N-1} \mathcal{E}_{l} = \frac{1}{N} \Tr \left[\mathcal{H}
    \right]
\end{equation}
consequently, given a specific circulant discrete structure, the ergotropy of any localized state is the same. 

\subsection{Complete Q-Cell}
Let us now study complete Q-Cell \cite{sachdev1993gapless}, as a second structure, which has an associated Hamiltonian of the form

\begin{equation}
\label{eq:Hamiltonian_complete}
    \mathcal{H}^{c} = -\mathcal{J}\sum_{\substack{m,l=0\\m > l}}^{N-1} \ket{x_{m}}\bra{x_{l}} + h.c.
\end{equation}
Such structure has only two distinct eigenvalues $\{ (1-N)\mathcal{J}, \mathcal{J} \}$, 
with one of them that is highly degenerate: 
\begin{equation}
    \label{eq:eigenvalues_complete_cell}
    \mathcal{E}^{c}_{l} = \{ (1-N)\mathcal{J} \; \text{deg=1} \; ; \; \mathcal{J} \; \text{deg=$N-1$} \} 
\end{equation}

\begin{figure}[!ht]
  \centering
  \includegraphics[width=0.3\columnwidth]{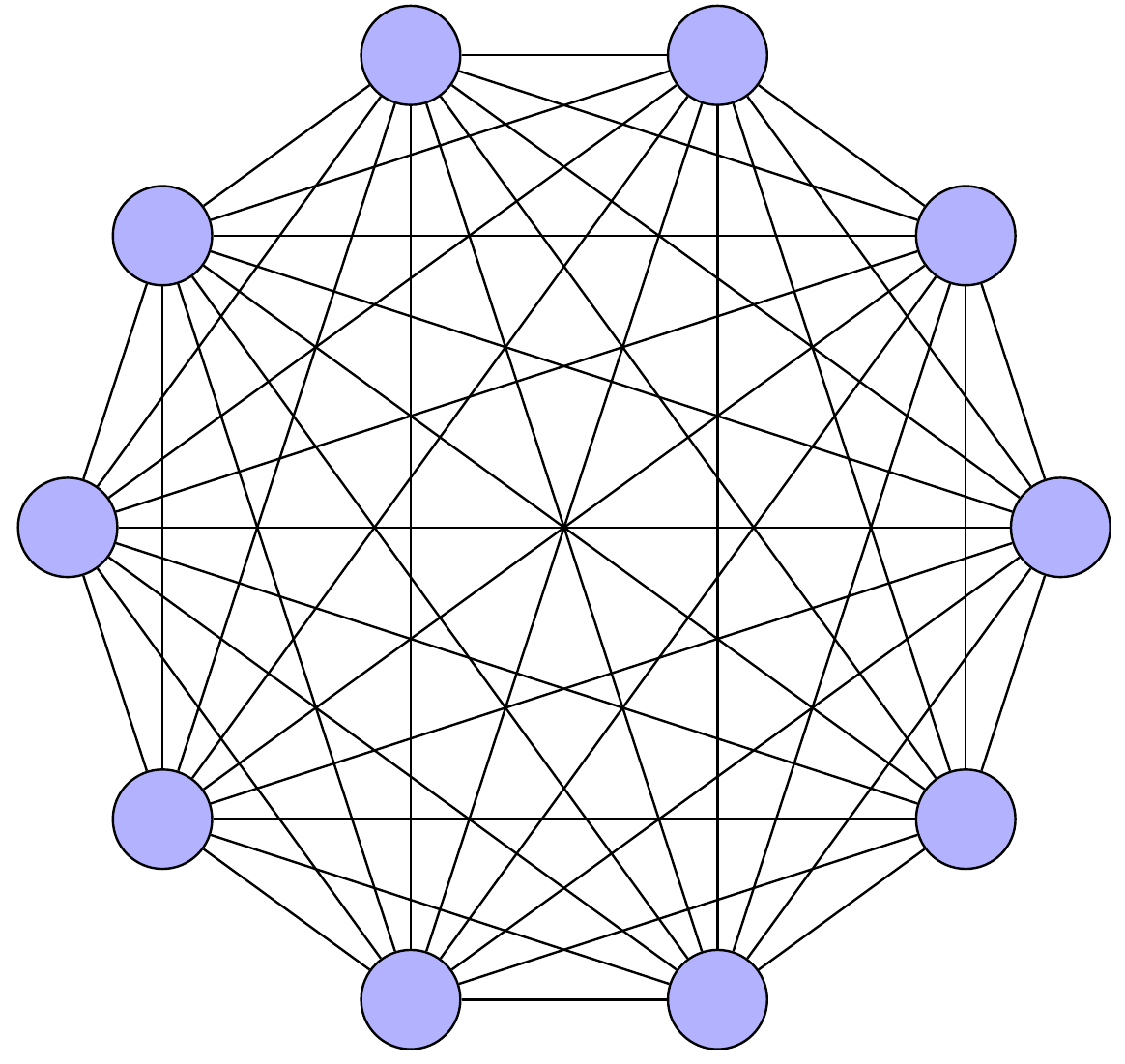}
    \caption{Graphical representation of the Complete Q-Cell composed of $N=10$ discrete elements for a circulant quantum battery.}
    \label{fig:Complete Battery}
\end{figure}

From Eq. \eqref{eq:max_ergotropy_pure_states} and the definition of the complete cell eigenvalues in Eq.\eqref{eq:eigenvalues_complete_cell}, the maximal work that can be extracted from a 
complete topology at fixed entropy is
\begin{equation}
    \mathcal{W}_{max}^{c} = N\mathcal{J},
\end{equation}
and is proportional to the size of the structure itself. To extract from the system the 
maximal amount of work at fixed entropy it is necessary that the initial state corresponds 
to any linear combination of eigenvectors associated to the eigenvalue $\mathcal{E}_{1}^{c}=\mathcal{J}$, and the final state after the application of the operator $\mathcal{U}$ 
must be the ground state of the system, i.e. the state associated to $\mathcal{E}_{0}^{c}=\mathcal{J}(1-N)$. A possible choice of an operator able to perform such transformation, 
has the form
\begin{equation}
    \label{eq:U_complete_cell}
    \mathcal{U} = \sum_{l=0}^{N-1} \ket{\phi_{l}}\bra{\phi_{N-1-l}}.
\end{equation}
Due to the degeneracies of eigenvalues different operator $\mathcal{U}$ can extract 
the same energy from the system leading to different final states.

As a paradigmatic example of non ideal initial state we consider again 
a localized state. In such scenario the ergotropy results
\begin{equation}
    \mathcal{W}_{loc}^{c} = (N-1)\mathcal{J} + \frac{1}{N} \Tr [\mathcal{H}^{c}]= (N-1)\mathcal{J},
\end{equation}
and, as for the ideal case, depends again on the dimensionality of the system. Then in such architecture, the higher is the dimensionality of the Q-Cell, the higher is the maximal ergotropy and the ergotropy associated to any localized state in the structure. For a localized state in the form $\ket{\psi_{0}}=\ket{x_{0}}$ a possible transformation for the maximal work extraction is
\begin{equation}
    \label{eq:U_matrix_complete_cell_loc}
    \mathcal{U} = \sum_{l=0}^{N-1} \ket{\phi_{l}}\bra{x_{l}}.
\end{equation}
As for the wheel Q-Cell, the localized state can be adopted as a convenient state to initialize the battery, before entering in the highest lowest eigenstate transition. At variance with the wheel battery, however, the ergotropy of localized states is comparable with the one associated to the highest energetic eigenstate: $\mathcal{W}^{c}_{max} - \mathcal{W}^{c}_{loc} = \mathcal{J}$.

The process of energy extraction from a localized state can be seen as the opposite of a search problem, since the initial state is a a specific vertex of the battery ($\ket{\psi_{0}}=\ket{x_{0}}$) and the final state is the ground state of the system, which corresponds to the equal superposition of all the vertices (see Eq.\eqref{eq:ground_state_circulant}). Consequently the time evolution operator adopted to perform such procedure may be inspired by such results \cite{childs2004spatial,frigerio2022quantum}. Using the dimensionality reduction method for a localized state, following Eqs.\eqref{eq:Taylor_Krylov}-\eqref{eq:Krylov_subspace}, the Krylov space \cite{razzoli2022universality} is the same as in Eq. \eqref{eq:Krylov_subspace_w_q-c}

\begin{equation}
\mathcal{I}(\mathcal{H}^{c},\ket{\psi_{0}}) = \{ \ket{\psi_{0}},\ket{s} \}
\label{eq:Krylov_subspace_c_u_c}
\end{equation}
where $\ket{s}$ is the equal superposition of all the discrete elements in the complete Q-Cell except $\ket{\psi_{0}}=\ket{x_{0}}$, i.e.,
\begin{equation}
\label{eq:ket_s}
\ket{s} = \frac{1}{\sqrt{N-1}} \sum_{l=1}^{N-1} \ket{x_l}.
\end{equation}
In such basis the Hamiltonian of the Q-Cell in Eq.\eqref{eq:Hamiltonian_complete} may be 
rewritten as
\begin{equation}
    \label{eq:Hamiltonian_complete_red}
    \mathcal{H}^{c}_{red} = -\mathcal{J}\left(
    \begin{array}{cc}
        0 & \sqrt{N-1} \\
        \sqrt{N-1} & N-2  
    \end{array}\right),
\end{equation}
and with the application of a localized potential $\mathcal{V}=-N\mathcal{J} \ket{\psi_{0}}\bra{\psi_{0}}$
in the system is induced the transition from the localized state to the ground state, as

\begin{equation}
    \label{eq:large cell limit method}
    e^{-i(\mathcal{H}^{c}_{red}+\mathcal{V})t^{*}}\ket{\psi_{0}} \simeq  \frac{1}{\sqrt{N}} \sum_{l=0}^{N-1} \ket{x_{l}} = \ket{\phi_{0}},
\end{equation}
and represent a suitable choice for work extraction for any localized state in the complete Q-Cell.

As we have done for the wheel Q-Cell, let us analyze the charge discharge cycle for a complete 
Q-cell in thermal equilibrium. In such condition the state of the battery is defined through a density matrix in the form of  Eq. (\ref{eq:thermal_rho}). Due to the degeneracies of the eigenvalues the charge procedure has to exchange the population of only two different energy levels. Accordingly, see Eq. (\ref{eq:ergotropy_thermal_rho}), when the system is in an inverse thermal state the ergotropy reads 
\begin{equation}
    \label{eq:ergotropy_thermal_complete}
    \mathcal{W}_{th,inv}^{c}= \frac{\mathcal{J}N\left(-e^{-\beta\mathcal{J}}+e^{-\beta(1-N)\mathcal{J}}\right)}{(N-1)e^{-\beta\mathcal{J}}+e^{-\beta(1-N)\mathcal{J}}}.
\end{equation}

\subsection{Ring Q-Cell}
\label{subsec:RUC}
As a second circulant discrete structure we study the ring topology, which has an associated Hamiltonian in the form 

\begin{equation}
\label{eq:ring_adjacency}
    \mathcal{H}^{r} = -\mathcal{J}\sum_{m=0}^{N-1} \ket{x_{m}} \bra{x_{m+1}} + h.c.
\end{equation}
with $\ket{x_{N}}=\ket{x_{0}}$, and has eigenvalues

\begin{equation}
    \mathcal{E}^{r}_{l} = -2 \mathcal{J}\cos \left( \frac{2\pi(l+1)}{N} \right) \; ; \; l = 0,...,N-1 
\end{equation}

\begin{figure}[!ht]
  \centering
  \includegraphics[width=0.3\columnwidth]{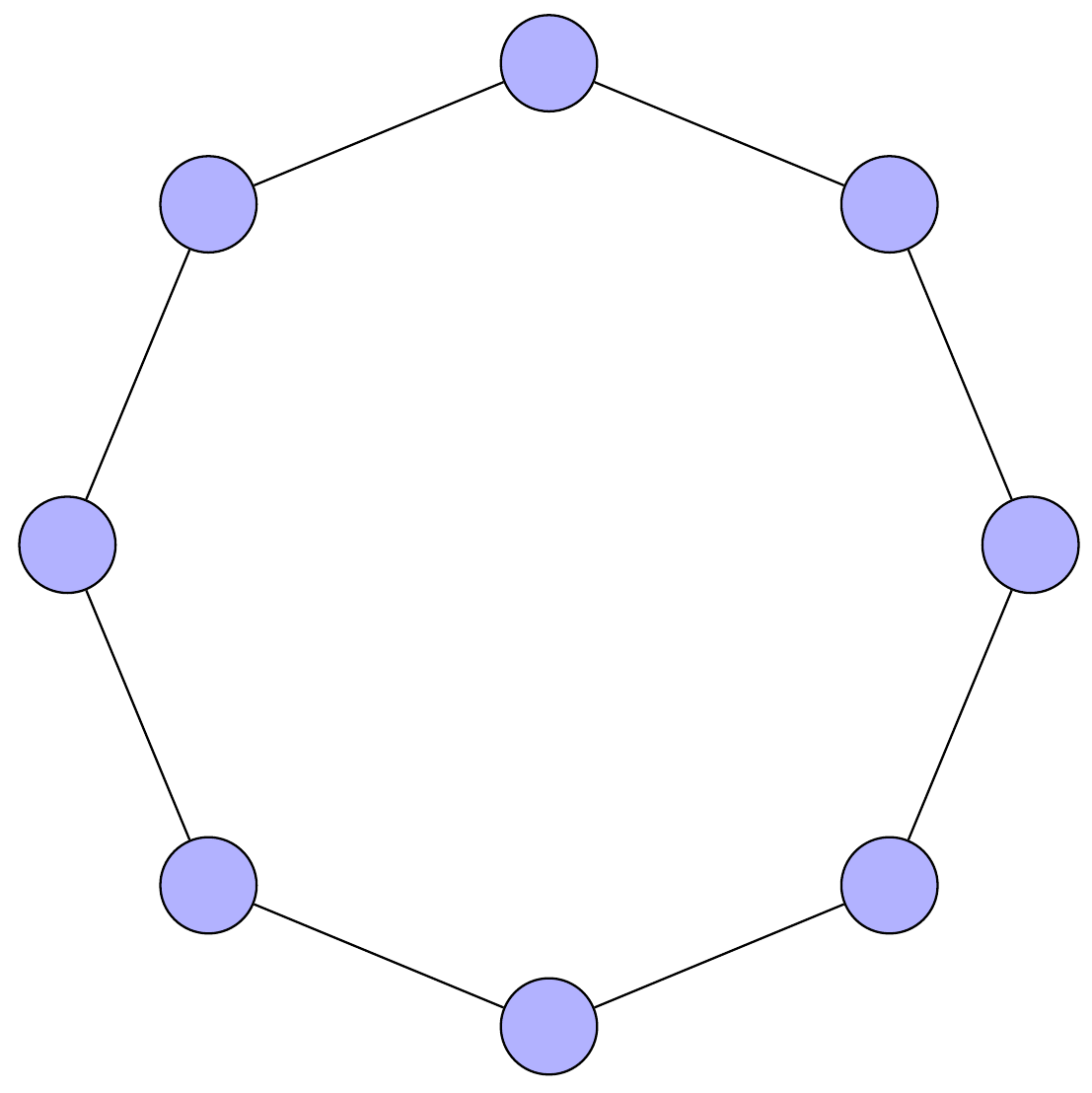}
    \caption{Graphical representation of the ring Q-Cell composed of $N=8$ discrete elements for a circulant quantum battery.}
    \label{fig:Ring Battery}
\end{figure}

The maximal work that can be extracted from a ring topology at fixed entropy is

\begin{equation}
    \mathcal{W}_{max}^{r} = \begin{cases}
         2\mathcal{J}\left( \cos\left(2\pi\right) - \cos\left(\pi\right)\right) = 4\mathcal{J} \;\;\;\; N=2K \\
         2\mathcal{J} - 2\mathcal{J} \cos \left( \pi \pm \frac{2\pi}{N} \right) < 4\mathcal{J} \;\;\;\;\;\;\;\; N=2K+1 \; ,
    \end{cases}
\end{equation}
This results shows that it is convenient to have a battery with an even number of elements with respect to a battery with the same topology and an odd number of discrete elements, independently of its dimension. 

The ergotropy is independent of the size of the system, at variance with complete cells where ergotropy growing linearly with size.  On the other hand, the practical challenges in constructing fully connected batteries must be taken into account. This consideration, combined with the size-independent ergotropy of ring-shaped batteries, led us to investigate the behavior of ring-based Q-Cells with 3 and 4 vertices more closely, particularly analyzing their performance under noise and decoherence.

For any localized state, as it happens for the two previous topologies (i.e. Wheel and Complete Q-Cells), the localized state can be adopted as a convenient state to 
initialize the battery. in a ring Q-Cell geometry the ergotropy reads

\begin{equation}
    \mathcal{W}_{loc}^{r} = 2\mathcal{J} - \frac{1}{N} \sum_{l}  2\mathcal{J}\cos \left( \frac{2\pi l}{N} \right) = 2\mathcal{J} + \frac{1}{N} Tr (\mathcal{H}^{r}) = 2\mathcal{J},
\end{equation}
Then, the dimensionality of the battery does not influence the maximal amount of energy which can be extracted from a localized state in a ring topology. In a physical implementation in which a state is initially localized in position space the most convenient set-up involves the minimal possible ring topology, i.e. a structure composed of only $N=3$ discrete elements. On the other hand, a structure with $N=4$ elements has the same associated ergotropy for a localized state, but also the maximal ergotropy achievable for a ring topology.

Moving to the analysis of non ideal scenario, for a ring Q-Cell in thermal equilibrium the ergotropy reads
{
\begin{equation}
    \label{eq:ergotropy_thermal_ring}
    \mathcal{W}_{th,inv}^{r}= \frac{ \textcolor{black}{ -2\mathcal{J} \sum_{l=0}^{N-1}  \left( \cos{\left(\frac{2\pi (\mathcal{S}+l)}{N} \right)} - \cos{\left(\frac{2\pi (N-l)}{N} \right)} \right) e^{2\mathcal{J}\beta\cos{\left(\frac{2\pi (N-l)}{N} \right)}} } }{ \sum_{l=0}^{N-1} e^{2\mathcal{J}\beta\cos{\left(\frac{2\pi (l+1)}{N} \right)}} }
\end{equation}}
where we recall $\mathcal{S}=(N-1)/2$ or $\mathcal{S}=N/2$ depending on the odd or even dimensionality of the cell.


The above results motivate a detailed analytical study of the $N=3,4$ cases, as they yield the best ergotropy results, while maintaining minimal the ring dimensionality, thereby minimizing resource cost of the Q-Cell.


\begin{figure}[!ht]
  \centering
  \includegraphics[width=0.3\columnwidth]{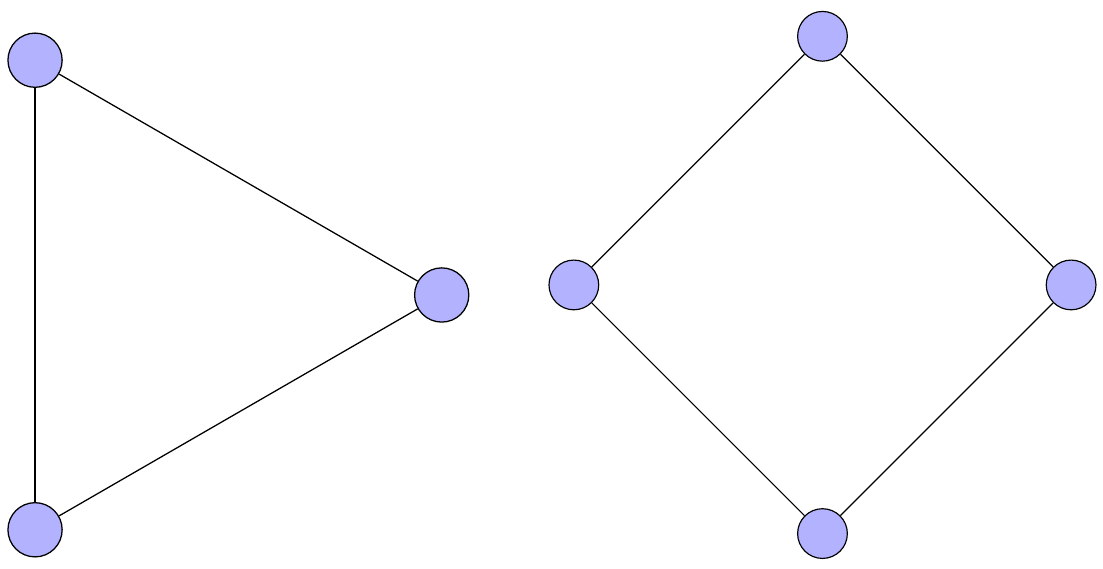}
    \caption{Minimal Q-Cells composed of $N=3,4$ discrete elements for a quantum battery. The results of Section \ref{subsec:RUC} motivates the study of such minimal structures because they maximize the ergotropy for a localized initial state for the ring structure minimizing the resources required to implement the battery.}
    \label{fig:minring}
\end{figure}

In the $N=3$ case the general Hamiltonian of the system expressed in Eq.\eqref{eq:ring_adjacency} in position space reads

\begin{equation} 
\label{eq:n3cell}
    \mathcal{H}^{r}_{N=3}=-\mathcal{J}
    \left(
    \begin{array}{ccc}
        0 & 1 & 1 \\
        1 & 0 & 1 \\
        1 & 1 & 0 
    \end{array}\right).
\end{equation}
Following the theory of circulant matrices the eigenvalues (in ascending order) and their corresponding eigenvectors are

\begin{align}
    \label{eq:eigenstate_n3cell}
    &\mathcal{E}^{r}_{0}=-2\mathcal{J}, \quad \ket{\phi_{0}} = \frac{1}{\sqrt{3}} \sum_{m=0}^{2} \ket{x_{m}} \nonumber \\
    &\mathcal{E}^{r}_{1}=+\mathcal{J}, \quad \ket{\phi_{1}} = \frac{1}{\sqrt{3}} \sum_{m=0}^{2} e^{\frac{i2m\pi}{3}}\ket{x_{m}}\nonumber \\
    &\mathcal{E}^{r}_{2}=+\mathcal{J}, \quad \ket{\phi_{2}} = \frac{1}{\sqrt{3}} \sum_{m=0}^{2} e^{\frac{i4m\pi}{3}}\ket{x_{m}}, 
\end{align}
with the system that is defined by only two distinct energy levels, due to the degeneracy $\mathcal{E}^{r}_{1}=\mathcal{E}^{r}_{2}=+\mathcal{J}$. 

To extract from the system the maximal amount of work at fixed entropy it is necessary that the initial state is $\ket{\psi_{0}}=\ket{\phi_{2}}$ (though, due to the degeneracy we could have chosen also a state $\ket{\psi_{0}}=\ket{\phi_{1}}$ or any superposition between $\ket{\phi_{1}}$ and $\ket{\phi_{2}}$; the transformation would differ only by a basis change term, but would lead to the same work extraction), and the final state after the application of the operator $\mathcal{U}$ must be the ground state of the system $\ket{\phi_{0}}$. A possible choice of an operator able to perform such transformation, has a structure in the form

\begin{equation}
    \label{eq:U_n3cell}
    \mathcal{U} = \ket{\phi_{2}}\bra{\phi_{0}} + \ket{\phi_{1}}\bra{\phi_{1}} + \ket{\phi_{0}}\bra{\phi_{2}}.
\end{equation}

For a localized state in the form $\ket{\psi_{0}}=\ket{x_{0}}$ a possible ideal transformation for the maximal work extraction is

\begin{equation}
    \label{eq:U_matrix_n3_loc}
    \mathcal{U} = \ket{\phi_{0}}\bra{x_{0}} + \ket{\phi_{1}}\bra{x_{1}} + \ket{\phi_{2}}\bra{x_{2}}.
\end{equation}

If we consider a thermal state of a ring of dimension $N=3$, Eq.\eqref{eq:ergotropy_thermal_ring} reduces to 

\begin{equation}
    \label{eq:ergotropy_thermal_ring_3}
    \mathcal{W}_{th,inv}^{r,N=3}= \frac{3\mathcal{J}\left( e^{2\beta\mathcal{J}}-e^{-\beta\mathcal{J}} \right)}{2e^{-\beta\mathcal{J}}+e^{2\beta\mathcal{J}}}
\end{equation}


In the $N=4$ case the Hamiltonian of the ring Q-Cell in position space is defined as

\begin{equation} 
\label{eq:n4cell}
    \mathcal{H}^{r}_{N=4}=-\mathcal{J}
    \left(
    \begin{array}{cccc}
         0 & 1 & 0 & 1 \\
        1 & 0 & 1 & 0 \\
         0 & 1 & 0 & 1 \\
        1 & 0 & 1 & 0
    \end{array}\right).
\end{equation}
Following the theory of circulant matrices the eigenvalues (in ascending order) and the corresponding eigenvectors are

\begin{align}
    \label{eq:eigenstate_n4cell}
    &\mathcal{E}^{r}_{0}=-2\mathcal{J}, \quad \ket{\phi_{0}} = \frac{1}{2} \sum_{m=0}^{3} \ket{x_{m}} \nonumber \\
    &\mathcal{E}^{r}_{1}=0, \quad \ket{\phi_{1}} = \frac{1}{2} \sum_{m=0}^{3} e^{\frac{im\pi}{2}}\ket{x_{m}}\nonumber \\
    &\mathcal{E}^{r}_{2}=0, \quad \ket{\phi_{2}} = \frac{1}{2} \sum_{m=0}^{3} e^{im\pi}\ket{x_{m}}\nonumber \\
    &\mathcal{E}^{r}_{3}=+2\mathcal{J}, \quad \ket{\phi_{3}} = \frac{1}{2} \sum_{m=0}^{3} e^{\frac{i3m\pi}{2}}\ket{x_{m}},
\end{align}
with the system that is defined by only three distinct energy levels,due to the degeneracy $\mathcal{E}^{r}_{1}=\mathcal{E}^{r}_{2}=0$.

As in the $N=3$ case, to extract from the system the maximal amount of work at fixed entropy it is necessary that the initial state is $\ket{\psi_{0}}=\ket{\phi_{3}}$, and the final state after the application of the operator $\mathcal{U}$ the final state must be the ground state of the system $\ket{\phi_{0}}$. A possible choice of an operator able to perform such transformation, has a structure in the form

\begin{equation}
    \label{eq:U_n4cell}
    \mathcal{U} = \ket{\phi_{3}}\bra{\phi_{0}} + \ket{\phi_{2}}\bra{\phi_{1}} + \ket{\phi_{1}}\bra{\phi_{2}}+\ket{\phi_{0}}\bra{\phi_{3}}.
\end{equation}
where the matrix $\mathcal{B}$ is now a four dimensional square matrix, which again follows the definition of Eq.\eqref{eq:projector}. This operator, together with the square ring Q-Cell, provides the ideal operation for work extraction in ring topologies.

For a localized state in the form $\ket{\psi_{0}}=\ket{x_{0}}$ a possible ideal transformation for the maximal work extraction is

\begin{equation}
    \label{eq:U_matrix_n4_loc}
    \mathcal{U} = \ket{\phi_{0}}\bra{x_{0}} + \ket{\phi_{1}}\bra{x_{1}} + \ket{\phi_{2}}\bra{x_{2}}+\ket{\phi_{3}}\bra{x_{3}}.
\end{equation}

If we consider a thermal state of a ring of dimension $N=4$, Eq.\eqref{eq:ergotropy_thermal_ring} reduces to 

\begin{equation}
    \label{eq:ergotropy_thermal_ring_4}
    \mathcal{W}_{th,inv}^{r,N=4}= \frac{4\mathcal{J}\sinh{\left( 2\beta\mathcal{J}\right)}}{1+\cosh{\left( 2\beta\mathcal{J}\right)}}
\end{equation}

\section{Noisy ring Q-Cells}
\label{sec:NC}

\textcolor{black}{The practical challenges in constructing large fully connected batteries must be taken into account. This consideration, combined with the size-independent ergotropy of ring-shaped cells, led us to investigate the behavior of ring-based Q-Cells with 3 and 4 discrete elements more closely, particularly analyzing their performance under noise and decoherence. Additionally, fixing the number of resources (i.e. discrete elements), the maximal ergotropy of a large complete Q-Cell correspond to the ergotropy of a collection of $N=3$ or $N=4$ Q-Cells.} Therefore, we describe here the non-ideal evolution of minimal ring cells, namely the triangle and the square of Fig. \ref{fig:minring}. In our model, the cell is prepared in a pure initial state  named charged state, and denoted in this section as $ {\rho}^c= {\rho}(0) = |\chi_0\rangle \langle \chi_0|$. Then, the cell is left untouched for a time $t$ in which the evolution of the state is affected by the environment, and is accordingly described by a decoherence model chosen among the three listed in Sec.\ref{sec:FQB}, as in Eq (\ref{eq:noisy_evolution}). 
Then, a unitary operation $\mathcal{U}_{d}$ is applied to the state ${\rho}(t)$ to extract as much energy as possible, performing the discharge of the cell. In order to extract the highest possible work (i.e. the ergotropy) from the state ${\rho}(t)$ one should know its spectral decomposition and perform the unitary in Eq (\ref{eq:ideal_transformation}). Thus, perfect knowledge of ${\rho}(t)$ is required, and this could not be the case due to the noisy character of the evolution. For this reason we considered three possible strategies: if one has complete knowledge about ${\rho}(t)$, then the highest possible work can be extracted using the unitary discharge $\mathcal U_{erg}$, given in Eq. (\ref{eq:ideal_transformation}). Otherwise, if one has no knowledge at all about the decoherence channels but still controls the free Hamiltonian $\mathcal{H}$, then he can optimize the discharge $\mathcal U_{free}$ on the state ${\rho}_{free}(t)$, which evolves accordingly to
\begin{equation}
\label{eq:noise_free}
    \partial_t {\rho}_{free}(t)=-i\left[ \mathcal{H},{\rho}_{free}(t) \right] .
\end{equation}
$\mathcal U_{free}$ is optimized using Eq (\ref{eq:ideal_transformation}) on the state ${\rho}_{free}(t)$. The last option is to apply the discharge
$\mathcal U_0$, optimized on the initial state ${\rho}(0)$. In the following, 
we study the three different decoherence models described in previous Sections, and compare the performance of the different discharges listed above.

\subsection{Pure dephasing}
\label{sec:NC_ID}
\begin{figure}[!ht]
    \centering
    \includegraphics[width=0.7\linewidth]{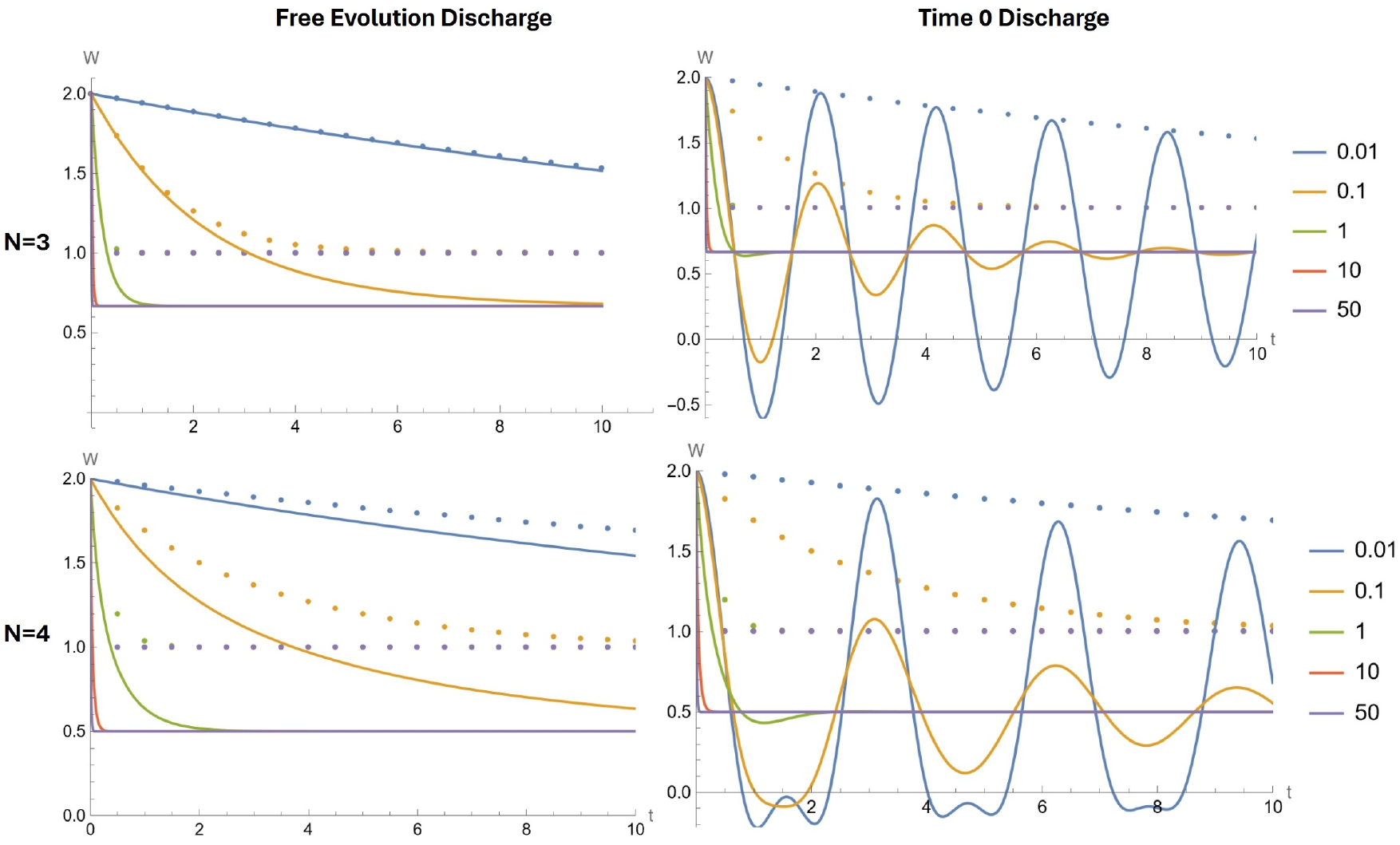}
    \caption{Work extraction under intrinsic decoherence for different unitary discharges, number of sites and noise parameters (depicted with different colors). The initial state is chosen as an arbitrary site of the ring. The free evolution discharge column represents the work extracted with $\mathcal U_{free}$ while the work extracted with $\mathcal U_0$ is in the second column. In all the plots, the dots represent the Ergotropy, i.e. the the work extracted using $\mathcal U_{erg}$. Note that the Ergotropy saturates at the value of 1 for both the triangle ($N=3$) and the square ($N=4$). This is because of the degenerate eigenvalues in their Hamiltonians.}
    \label{fig:id_site}
\end{figure}

We consider here a QW subject to decoherence in the energy basis, evolving as in Eq. (\ref{eq:decoherence_energy_basis}). We analyze two different types of charged states: the highest energy eigenstate or a localized state.\\
Let's start with a state initialized in the highest energy state, which is the best scenario in order to increase the extractable energy. The noisy evolution is easily obtained projecting Eq.(\ref{eq:decoherence_energy_basis}) in the energy basis: $\{|\phi_{\mu}\rangle\}:\mathcal{H}|\phi_{\mu}\rangle=\mathcal{E}_{\mu}|\phi_{\mu}\rangle$. We obtain
\begin{equation}
    \rho_{\mu\nu}(t)=\rho_{\mu\nu}(0)e^{-i(\mathcal E_{\mu}-\mathcal E_{\nu})t-\frac{\gamma}2(\mathcal E_{\mu}-\mathcal E_{\nu})^2t},
\end{equation}
which leads to an exponential decay the off-diagonal terms. For this reason the initial state (being an energy eigenstate) is left unchanged by this type of evolution, and the ergotropy stays the initial one, namely $\mathcal E_{max}-\mathcal E_{min}$.\\
A state initialized in a specific site of the ring, let's say ${\rho}^c=|x_j\rangle\langle x_j|$, behaves differently. We can write $|x_j\rangle$ in the energy basis as
\begin{equation}
    |x_j\rangle=\frac1{\sqrt N}\sum_{\mu=0}^{N-1} \omega^{-j\mu}|\phi_{\mu}\rangle, \quad \mbox{with}\, \omega=e^{\frac{2i\pi}{N}}.
\end{equation}
and
\begin{equation}
    \rho^c=|x_j\rangle\langle x_j|=\frac1{N}\sum_{\alpha=0}^{N-1} \sum_{\beta=0}^{N-1}\omega^{-j\alpha}\omega^{j\beta}|\phi_{\alpha}\rangle \langle \phi_{\beta}|.
\end{equation}
So
\begin{equation}
    \rho^c_{\mu\nu}=\langle\phi_{\mu}|\rho^c|\phi_{\nu}\rangle=\frac1{N}\omega^{j(\nu-\mu)}.
\end{equation}
and the evolution is:
\begin{equation}
    \rho_{\mu\nu}(t)=\frac1{N}\omega^{j(\nu-\mu)}e^{-i(\mathcal E_{\mu}-\mathcal E_{\nu})t-\frac{\gamma}2(\mathcal E_{\mu}-\mathcal E_{\nu})^2t}.
\end{equation}
If the Hamiltonian is non degenerate, then the final state is passive and diagonal in the energy basis:
\begin{equation}
    {\rho}(t \rightarrow \infty)=\frac1N\sum_{\mu}|\phi_{\mu}\rangle\langle\phi_{\mu}|,
\end{equation}
 from which there is no way to extract energy with a unitary operation.\\
If instead the Hamiltonian has degenerate eigenvalues, certain sectors of the density matrix remain protected from decoherence. Consequently, as time approaches infinity, the density matrix in the energy basis evolves into a block-diagonal form, where the block sizes correspond to the degrees of degeneracy. This configuration enables the system to preserve some extractable energy indefinitely, ensuring that the ergotropy does not go to zero. 
In particular, for ring cells with $N=3,4$ we recall that we have 
\begin{equation}
    \mathcal{H}^{r}_{N=3}= -\mathcal{J} \left( \ket{0} \bra{1} + \ket{1} \bra{2} + \ket{0} \bra{2} + h.c. \right) \quad \mbox{with eigenvalues: }(\mathcal{J},\mathcal{J},-2\mathcal{J})
\end{equation}
and
\begin{equation}
   \mathcal{H}^{r}_{N=4}=-\mathcal{J} \left( \ket{0} \bra{1} + \ket{1} \bra{2} + \ket{2} \bra{3} + \ket{0} \bra{3} + h.c. \right) \quad \mbox{with eigenvalues: }(2\mathcal{J},0,0,-2\mathcal{J}).
\end{equation}
Starting from a localized state, the presence of degenerate eigenvalues keeps the 
ergotropy from going to zero in the long time limit, as we can see in Fig. \ref{fig:id_site}.

\subsection{Haken-Strobl decoherence}
\label{sec:NC_HS}
In this subsection we look at the evolution performed by the Haken-Strobl decoherence model, in which the jump operators are the projectors on the sites of the ring. Again, we could initialize the state in the highest energy eigenstate or in a certain site. In both cases we let the system evolve for a time $t$, and then we perform a unitary discharge, chosen among the three listed before: $\mathcal U_{erg}, \mathcal U_{free}$ or $\mathcal U_0$.

\begin{figure}[!ht]
    \centering
    \includegraphics[width=0.7\linewidth]{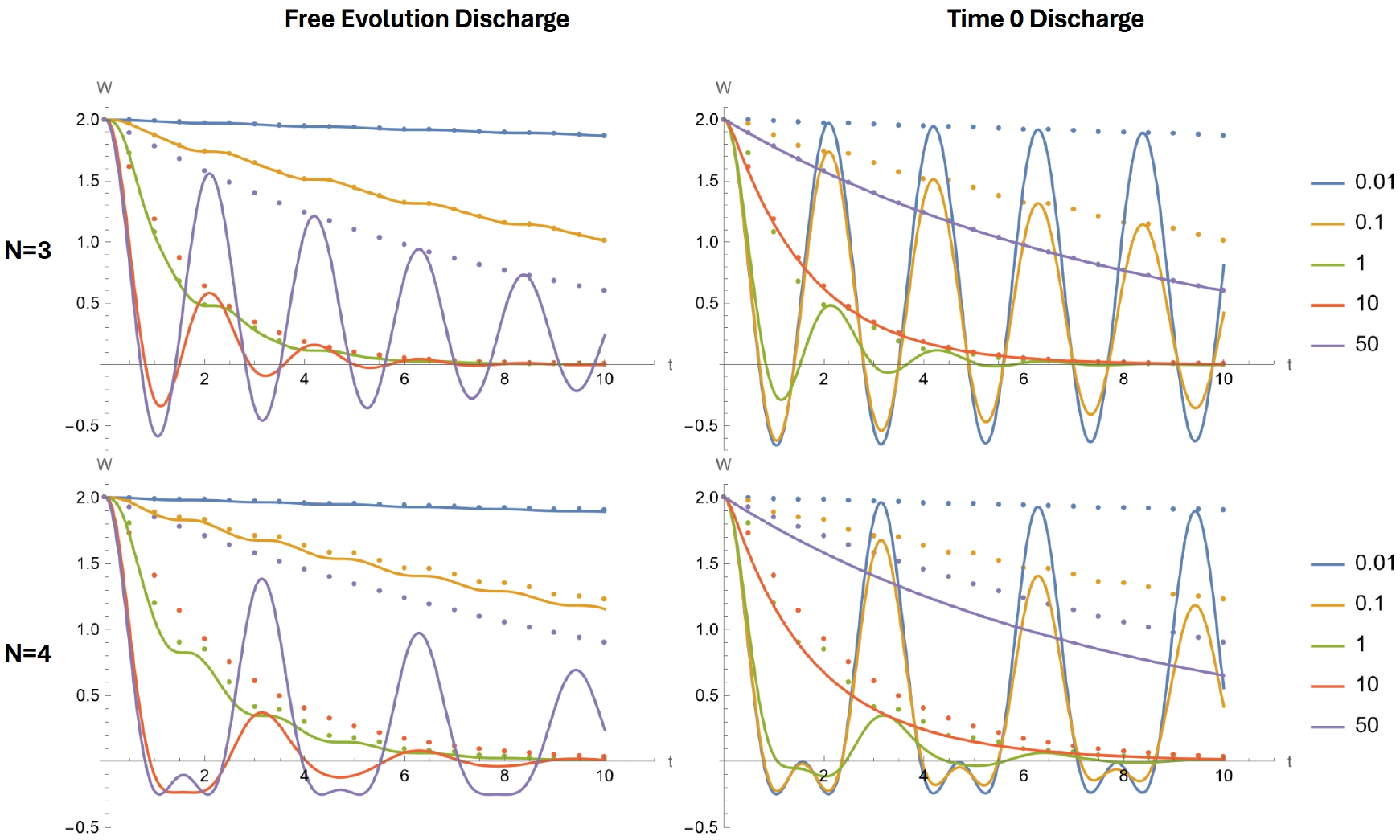}
    \caption{Work extraction under Haken-Strobl decoherence for different unitary discharges, number of sites and noise parameters (depicted with different colors). The initial state is chosen as an arbitrary site of the ring. The free evolution discharge column represents the work extracted with $\mathcal U_{free}$ while the work extracted with $\mathcal U_0$ is in the second column. In all the plots, the dots represent the Ergotropy, i.e. the the work extracted using $\mathcal U_{erg}$. Note that for high values of the noise parameter the line obtained with $\mathcal U_0$ is closer to the ergotropy while for small values of the noise parameter $\mathcal U_{free}$ performs better.
    }
    \label{fig:hs_site}
\end{figure}

In Fig. \ref{fig:hs_site} we plot with continuous lines (each color corresponding to a different value of the noise parameter) the work extracted performing a unitary discharge $\mathcal U_{free}$ (left column) or $\mathcal U_0$ (right column) at time $t$, starting from a localized state. The dotted lines represent instead the Ergotropy, obtained performing $\mathcal U_{erg}$. We see that for small values of the noise parameter the discharge optimized on the free evolution $\mathcal U_{free}$ performs better than the one optimized at time $t=0$ (except for some points in which the Ergotropy is very close to the work extracted with $\mathcal U_{0}$, caused by the oscillations of the latter). In fact, in this case, the full evolution is well approximated by the noiseless evolution of  Eq. (\ref{eq:noise_free}). For high values of the noise parameter, instead, $\mathcal U_0$ performs better than $\mathcal U_{free}$. We can explain this with a sort of Zeno effect, because the jump operators are projectors to the sites and for high noise intensity the evolution is frozen in the initial state.

\begin{figure}[!ht]
    \centering
    \includegraphics[width=0.7\linewidth]{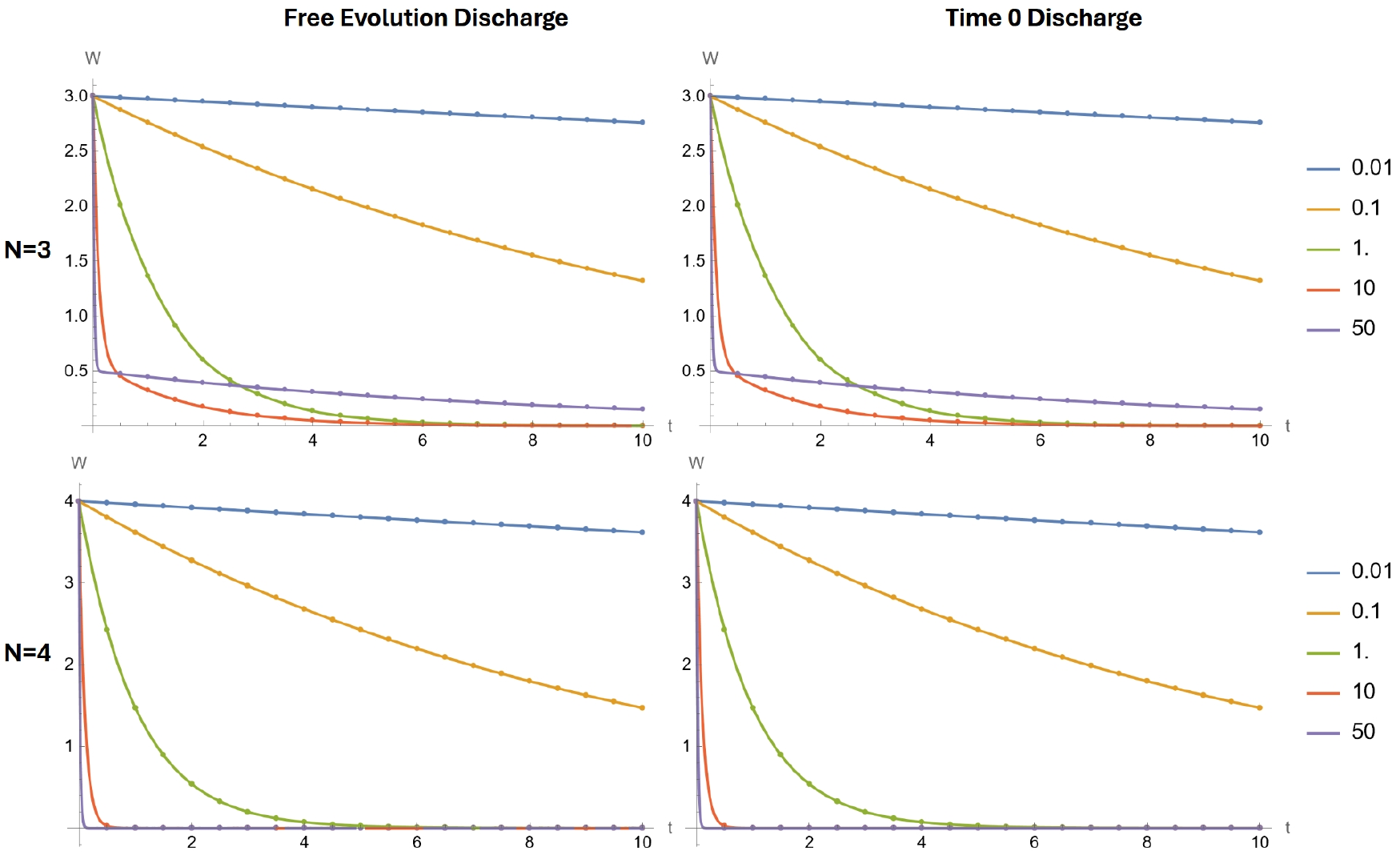}
    \caption{Work extraction under Haken-Strobl decoherence for different unitary discharges, number of sites and noise parameters (depicted with different colors). The initial state is chosen as the highest energy eigenstate. The free evolution discharge column represents the work extracted with $\mathcal U_{free}$ while the work extracted with $\mathcal U_0$ is in the second column. In all the plots, the dots represent the Ergotropy, i.e. the the work extracted using $\mathcal U_{erg}$. Note that they all coincide.}
    \label{fig:hs_energy}
\end{figure}

In Fig. \ref{fig:hs_energy} we show results of the same analysis, but starting from the highest energy eigenstate. As it is apparent from the plot, there is no difference in the work extracted with the three unitary discharges. The work extracted with $\mathcal U_0$ and $\mathcal U_{free}$ is equal simply because of the trivial evolution performed by the free Hamiltonian on the charged state. However, these two also coincide with the ergotropy. This is true because of the symmetries of the initial state compared to the noise model. In particular, this is valid because all the jump operators were equally weighted.  If this is the case, the unitary discharge saturating the ergotropy is the one optimized for the initial pure state throughout all the evolution. More details are  provided in Appendix \ref{sec:app_NC_HS}.

\subsection{Stochastic Quantum Walks}
\label{sec:NC_SQW}
The third noise model we consider is the one obtained  by the combination of the standard coherent evolution with weight $(1-p)$ and a classical random walk evolution with weight $p$, as shown in shown in Eq.(\ref{eq:quantum_stochastic_walk}). We performed the same analysis done for the Haken-Strobl decoherence model obtaining very similar results. The only difference is that, for high values of the noise parameter and localized initial state, we do not see that Zeno-like feature because the jump operators are not the projectors on the sites. Still, we see that using the highest energy eigenstate as charged state, the works extracted with ${\mathcal U}_{erg}$, ${\mathcal U}_{free}$ and $\mathcal U_0$ all coincide. Given similarity we put the results obtained for the triangle and square ring cells, in Appendix \ref{sec:app_NC_SQW}.

\section{Chirality as Ergotropy Enhancer}
\label{sec:CaEE}
\textcolor{black}{Due to the scaling of the ergotropy, the best battery performance is achieved by the ring- and complete-based Q-cells. These theoretical results, together with the noise resilience observed in the $N = 3,4$ ring cells, motivate a broader investigation of quantum effects in such circulant structures.} Indeed, the chirality of the system, lead to remarkable thermodynamic advantages. For a chiral circulant adjacency matrix, the eigenvectors are still provided by Eq. \eqref{eq:eigenvectors_circulant}, while the eigenvalues can be written as
\begin{align}
    \mathcal{E}_j&= 2 \Re\{q_1 \omega^j+q_2 \omega^{2j}+...+q_{\frac{n-1}2}\omega^{\frac{(n-1)j}2}\} \quad \mbox{if $n$ odd} \nonumber \\
    \mathcal{E}_j&= q_{\frac n2} (-1)^j+2 \Re\{q_1 \omega^j+q_2 \omega^{2j}+...+q_{\frac{n-2}2}\omega^{\frac{(n-2)j}2}\} \quad \mbox{if $n$ even}
\end{align}
where $q_{n-j}=q_j^*$ ($q_0$ in Eq.\eqref{eq:eigenvalues_circulant} is set to $0$ since we use the adjacency matrix). 

\subsection{Chiral Ring Q-Cell}

For a ring Q-Cell we have mathematically demonstrated that there are different $\Delta^{r}$ depending only on even or odd number of elements in the Q-Cell. The presence of complex elements in the chiral adjacency can promote the thermodynamic performances of a odd to the ones of an even Q-Cell. Considering a general ring Hamiltonian in the form

\begin{equation}
    \label{eq:chiral_ring}
    \mathcal{H}^{r,ch} = -\mathcal{J}\sum_{m=0}^{N-1} e^{i\gamma}\ket{x_{m}} \bra{x_{m+1}} + h.c.
\end{equation}
with, again, $\ket{x_{N-1}}=\ket{x_{0}}$.Then, the eigenvalues read
\begin{equation}
    \mathcal{E}^{r,ch}_l(\gamma)=-2\mathcal{J}\cos\left(\gamma+\frac{2\pi l}{N}\right) \; ; \; l = 1,...,N.
\end{equation}
The difference between two eigenvalues is then
\begin{align}
    \mathcal{E}^{r,ch}_{\mu}(\gamma)-\mathcal{E}^{r,ch}_{\nu}(\gamma)&=-2\mathcal{J}\left[\cos\left(\gamma+\frac{2\pi \mu}{N}\right)-\cos\left(\gamma+\frac{2\pi \nu}{N}\right)\right] \nonumber \\
    &=4\mathcal{J}\sin\left(\gamma+\frac{\pi}{N}(\mu+\nu)\right)\sin\left(\frac{\pi}{N}(\mu-\nu)\right)
\end{align}

If $N=2K+1$, it is maximized for $\gamma=\frac{\pi}2-\frac{\pi}{N}(\mu+\nu)$ and choosing $\mu-\nu=-\frac{N-1}2$, resulting in
\begin{align}
\label{eq:delta_ring_chiral_odd}
    &\mathcal{W}^{r,N=2K+1}_{max,ch}=\max_{\gamma,\mu,\nu}^{N=2K+1}\{\mathcal{E}^{r,ch}_{\mu}(\gamma)-\mathcal{E}^{r,ch}_{\nu}(\gamma)\}=4\mathcal{J}\sin\left(\frac{\pi (N-1)}{2 N}\right).
\end{align}
This means the the presence of ergotropy can enlarge the energy gap for a ring Q-Cell, leading to a larger ergotropy. 

On the other hand, if $N=2K$ is even, instead, we can choose $\mu-\nu=-\frac{N}2$, and saturate to the value of $4$ choosing $\gamma=0$

\begin{equation}
\label{eq:delta_ring_chiral_even}
    \mathcal{W}^{r,N=2K}_{max,ch}(\gamma)=\max_{\gamma,\mu,\nu}^{N=2K}\{\mathcal{E}^{r,ch}_{\mu}(\gamma)-\mathcal{E}^{r,ch}_{\nu}(\gamma)\}=4\mathcal{J}.
\end{equation}

For all the dimensionality (even and odd number of discrete elements in the chiral cell), the ergotropy for a localized state remains unchanged as the energy of both the ground state and of a localized state does not change.

\subsection{Chiral Complete Q-Cell}

The largest energy gap we found in our model is associated to the Complete Q-Cell, as $\Delta^{c}=N\mathcal{J}$. This directly derive from the possible values that the eigenvalue of a circulant matrix can assume. Starting from the definition of the circulant eigenvalues in Eq.\eqref{eq:eigenvalues_circulant} the maximal and minimal eigenvalue associated to a standard circular Hamiltonian matrix derived from the adjacency of a complete Q-Cell are respectively

\begin{align}
    \label{eq:max_min_eigenvalues_circulant}
    & max\left( -\sum_{m=0}^{N-1} \mathcal{J}  \omega^{m l} \right) = \mathcal{J} \nonumber \\
    & min\left( -\sum_{m=0}^{N-1} \mathcal{J}  \omega^{m l} \right) = -(N-1)\mathcal{J}
\end{align}
Considering a general complete chiral Q-Cell Hamiltonian in the form

\begin{equation}
    \label{eq:chiral_complete}
    \mathcal{H}^{c,ch} = -\mathcal{J}\sum_{\substack{m,l=0 \\ m>l}}^{N-1}e^{i\gamma_{m,l}}\ket{x_{m}} \bra{x_{l}} + h.c.
\end{equation}
with the condition on the phases such that $\gamma_{m,l} = -\gamma_{l,m}$ and $\gamma_{m,l} = \gamma_{m+1,l+1}$ (with the periodic boundary conditions $\gamma_{m,N}=\gamma_{m,0}$ and $\gamma_{N,l}=\gamma_{0,l}$) to preserve the circulant structure of the matrix. 

The eigenvalues of a general circulant complete chiral adjacency matrix in the form Eq.\eqref{eq:chiral_complete} for an odd number of elements is
\begin{equation}
    \mathcal{E}_l=2 \Re\{\sum_{m=1}^{\frac{N-1}{2}} q_m \omega^{lm}\}=-2\mathcal{J} \sum_{m=1}^{\frac{N-1}{2}}\cos\left(\gamma_{m,l}+\frac{2\pi l m}{n}\right).
\end{equation}
So, the difference between two of them is
\begin{align}
    \mathcal{E}_{\mu}-\mathcal{E}_{\nu}&=2\mathcal{J}\sum_{m=1}^{\frac{N-1}{2}}\left[\cos\left(\gamma_{m,l}+\frac{2\pi m\mu}{N}\right)-\cos\left(\gamma_{m,l}+\frac{2\pi m \nu}{N}\right)\right] \nonumber \\
    &=-4\mathcal{J}\sum_{m=1}^{\frac{N-1}{2}}\left[\sin\left(\gamma_{m,l}+\frac{\pi m}{N}(\mu+\nu)\right)\sin\left(\frac{\pi m}{N}(\mu-\nu)\right)\right]
\end{align}
 The absolute value of the expression is maximized choosing $\gamma_{m,l}=\frac{\pi}2-\frac{\pi m}{N}(\mu+\nu)$ and $\mu-\nu=$ either $1$ or $2$ . In this case the difference would be
\begin{align}
    \mathcal{W}^{c,N=2K+1}_{max,ch}=&\max_{\gamma_{m,l},\mu,\nu}^{N=2K+1}\{\mathcal{E}_{\mu}-\mathcal{E}_{\nu}\}=-4\mathcal{J}\sum_{m=1}^{\frac{N-1}{2}}\sin\left(\frac{\pi m}{ N}\right) \nonumber \\
    &=-2\mathcal{J}\csc{\left( \frac{\pi}{2N} \right)}\sin{\left( \frac{(N-1)\pi}{2N}\right)}.
\end{align}

The eigenvalues of a general circulant complete chiral adjacency matrix in the form Eq.\eqref{eq:chiral_complete} for an even number of elements is 
\begin{equation}
    \mathcal{E}_l=q_{\frac{N}{2}} \omega^{l \frac{N}{2}}+ 2 \Re\{\sum_{m=1}^{\frac{N-2}{2}} q_m \omega^{lm}\}= \mathcal{J}(-1)^l + 2 \mathcal{J} \sum_{m=1}^{\frac{N-2}{2}}\cos\left(\gamma_{m,l}+\frac{2\pi ml}{N}\right),
\end{equation}
where hermiticity forces $q_{\frac{N}{2}}=\pm\mathcal{J}$.
So, the difference between two of them is
\begin{align}
    &\mathcal{E}_{\mu}-\mathcal{E}_{\nu}=  \nonumber \\
    &q_{\frac{N}{2}}\left((-1)^{\mu}-(-1)^{\nu}\right)+2\mathcal{J}\sum_{m=1}^{\frac{N-2}{2}}\left[\cos\left(\gamma_{m,l}+\frac{2\pi m\mu}{N}\right)-\cos\left(\gamma_{m,l}+\frac{2\pi m \nu}{N}\right)\right] \nonumber \\
    &=q_{\frac{N}{2}}\left((-1)^{\mu}-(-1)^{\nu}\right)-4\mathcal{J}\sum_{m=1}^{\frac{N-2}{2}}\left[\sin\left(\gamma_{m,l}+\frac{\pi m}{N}(\mu+\nu)\right)\sin\left(\frac{\pi m}{N}(\mu-\nu)\right)\right].
\end{align}
The absolute value of the expression is maximized by choosing $\gamma_{m,l}=\frac{\pi}2-\frac{\pi m}{N}(\mu+\nu)$, $\mu-\nu=1$ and $q_{\frac{N}{2}} (-1)^{\nu}=1$.
\begin{align}
    \mathcal{W}^{c,N=2K}_{max,ch}=\max_{\gamma_{m,l},\mu,\nu}^{N=2K}\{\mathcal{E}_{\mu}-\mathcal{E}_{\nu}\}&=-2\mathcal{J}-4\mathcal{J}\sum_{m=1}^{\frac{N-2}{2}}\sin\left(\frac{\pi m}{ N}\right) \nonumber \\
    &=-2\mathcal{J}\cot{\left(\frac{\pi}{2N}\right)}
\end{align}
For all the dimensionality (even and odd number of discrete elements in the chiral cell), the ergotropy for a localized state remains unchanged as the energy of both the ground state and of a localized state does not change.

\section{Conclusions}
\label{sec:C}
In this work, we have employed the quantum walk formalism to analyze the behavior of quantum batteries composed of Q-cell units, illustrating how their structural and chiral properties govern performance. Three architectures, the ring, complete, and wheel configurations, have been systematically evaluated, with their ergotropy and unitary work extraction protocols derived explicitly from the underlying quantum walk dynamics. 

Our findings reveal distinct scaling behavior of ergotropy with system size $N$ for each architecture. Complete cells exhibit optimal performance, with ergotropy scaling linearly with $N$, while wheel configurations follow a $\sim \sqrt{N}$ scaling, and ring configurations display size-independent ergotropy. Given the experimental challenges in realizing fully connected complete cells, we further analyzed minimal ring cells ($N=3$ and $N=4$) as practical alternatives.  

Chirality was shown to play a relevant role in these systems. In complete graphs, it enhances ergotropy without altering the linear scaling, whereas in ring cells, it bridges the performance gap between odd and even $N$ configurations—with even-$N$ cells proving superior when no phase is applied in the Hamiltonian. 

The quantum walk framework also enabled the detailed investigation of noise resilience in ring cells, demonstrating how decoherence and thermalization impact performance. Three distinct decoherence models were analyzed, revealing how their effects on ergotropy depend on both the initial charged state and the unitary discharge protocol. Notably, chirality emerges as a resource to enforce Hamiltonian degeneracies, thereby protecting ergotropy from vanishing under intrinsic decoherence. 

These results establish fundamental design principles for quantum batteries, highlighting topology and chiral control as critical resources for optimizing energy storage and extraction. By linking the spectral properties of the CTQW Hamiltonian to extractable work, our analysis provides concrete guidelines for engineering efficient quantum batteries. This approach opens new avenues for developing scalable quantum energy devices that harness coherent dynamics, offering a pathway toward practical quantum-enhanced energy technologies. Future work will be about extending the analysis to non‑Markovian environments, exploring optimal control beyond step functions, and integrating measurement‑based feedback.

\appendix
\section{Additional details on noisy cells}
\label{sec:app_NC}
In this section we insert additional details about the extractable work from noisy cells that were omitted in Section \ref{sec:NC} .
\subsection{Haken-Strobl noisy dynamics for ring cells prepared in the highest energy eigenstate}
\label{sec:app_NC_HS}
We provide here additional details on the work extraction in ring cells affected by Haken-Strobl decoherence, when the charged state chosen is the highest eigenstate of the Hamiltonian.

As explained in Section \ref{sec:NC_HS}, we prepare the cell in the highest energy eigenstate, referred to as charged state. Then we let the system interact with the environment for a time $t$, in which we assume that the evolution is described by
\begin{equation}
    \label{eq:haken_strobl_detailed}
    \partial_{t} {\rho}(t) = -i\left[ \mathcal{H},{\rho}(t) \right] + \gamma \sum_{k} \left[P_{k}\, {\rho}(t)\,P_k-\frac12\left(P_{k}\, {\rho}(t)+{\rho}(t)\,P_k\right)\right],
\end{equation}
where $\gamma$ is a parameter controlling the strength of the decoherence, $\mathcal{H}$ is the "free Hamiltonian" and $P_k=|x_k\rangle\langle x_k|$ is the projector on the $k^{th}$ site.
The eigenvectors of the Hamiltonian are
\begin{equation}
    \ket{\phi_{\xi}} = \frac{1}{\sqrt{N}} \sum_{m} \omega^{\xi\,m} \ket{x_{m}} \quad 0\leq \xi,m\leq N-1
\end{equation}
with $\omega=\exp{\frac{2i\pi}{N}}$ and $\{|x_m\rangle\}$ being the $N-$dimensional site basis. Please note that in this section, to increase the readability, we use Latin letters to label the site basis and Greek letters to label the energy basis. The corresponding eigenvalues are
\begin{equation}
    \mathcal E_{\xi}= 2 \cos{\left(\varphi+\frac{2\pi \xi}{N}\right)},
\end{equation}
where we assumed the presence of a possible chiral phase $\varphi$. 
If we choose an eigenstate of the Hamiltonian as the charged state, the corresponding density matrix has equal diagonal elements in the site basis:
\begin{align}
    \rho^c_{kk}&\equiv\langle x_k|{\rho}^c|x_k\rangle\\
    &=\frac1{\sqrt{N}} \sum_{\mu}\omega^{k \mu}\langle\phi_{\mu}|\phi_{\xi}\rangle\langle\phi_{\xi}|\frac1{\sqrt N}\sum_{\nu}\omega^{-k\nu}|\phi_{\nu}\rangle\\
    &=\frac1N.
\end{align}
This feature allows to analytically solve eq.(\ref{eq:haken_strobl_detailed}), which can be written as
\begin{align}
\label{eq:hs_general_siteev}
    \partial_{t} {\rho}(t) &= -i\left[ \mathcal{H},{\rho}(t) \right] + \gamma \sum_{k} \rho_{kk}(t)\,|x_k\rangle\langle x_k|- \gamma \,{\rho}(t)\\
    &=-i\left[ \mathcal{H},{\rho}(t) \right] + \gamma\left(\mathbb J-\mathbb I\right)\circ{\rho}(t),
    \label{eq:hs_general}
\end{align}
where $\mathbb J$ is a matrix with all ones, $\mathbb I$ is the identity matrix and $(\star\,\circ\,\star)$ denotes the element-wise product between matrices. This means that $\left(\mathbb J-\mathbb I\right)\circ{\rho}(t)$ is the density matrix without the diagonal part. Let us now show that the diagonal part of ${\rho}(t)$ in the site basis does not evolve. Starting from Eq.(\ref{eq:hs_general}) we have
\begin{equation}
    \begin{cases}
      \partial_t\left(\mathbb I\circ{\rho}(t)\right)=-i\,\mathbb I\circ\left[ \mathcal{H},{\rho}(t) \right]   \\
      \partial_t \left[\mathcal H,{\rho}(t)\right]=-i\left[\mathcal H,\left[\mathcal H,{\rho}(t)\right]\right]-\gamma \left[\mathcal H,{\rho}(t)\right]+\gamma\left[\mathcal H,\mathbb I\circ {\rho}(t)\right].
    \end{cases}\,.
\end{equation}
Since at $t=0$ we have
\begin{equation}
\left\{
\begin{array}{c}
    \mathbb I\circ{\rho}^c=\frac 1N \mathbb I  \\
      \left[\mathcal H,{\rho}^c\right]=0,
\end{array}
    \right.
\end{equation}
then
\begin{equation}
\left\{
\begin{array}{c}
    \mathbb I\circ{\rho}(t)=\frac 1N \mathbb I  \\
      \left[\mathcal H,{\rho}(t)\right]=0
\end{array}
    \right. \quad \forall t,
    \label{eq:hs_noevol}
\end{equation}
meaning that the density matrix remains diagonal in the energy basis, and its diagonal part in the site basis remains proportional to the identity. We can now investigate how the eigenvalues of the density matrix change over time.
\\
Projecting eq.(\ref{eq:hs_general_siteev}) into the energy basis $\{|\phi_{\xi}\rangle\}$, and knowing that the diagonal elements of the density matrix in the site basis do not evolve: $\langle x_k|{\rho}(t)|x_k\rangle=\rho_{kk}(t)=\frac1N\, \forall t$,  we have
\begin{align}
    \partial_{t} \rho_{\mu \nu}(t) &= \partial_{t} \langle \phi_{\mu}|{\rho}(t)|\phi_{\nu}\rangle \\
    &= -i\left(\mathcal{E}_{\mu}-\mathcal{E}_{\nu}\right) \rho_{\mu \nu}(t) -\gamma\, \rho_{\mu \nu}(t) + \gamma \sum_{k} \rho_{kk}(t)\,\langle\phi_{\mu}|x_k\rangle\langle x_k|\phi_{\nu}\rangle\\
    &= i\left(\mathcal{E}_{\mu}-\mathcal{E}_{\nu}\right) \rho_{\mu \nu}(t) -\gamma\, \rho_{\mu \nu}(t) + \frac{\gamma}{N} \delta_{\mu \nu}.
\end{align}
Since the charged state is chosen as the highest energy eigenstate, let's say $|\phi_{\alpha}\rangle$, then in the energy basis $\rho^c_{\alpha \alpha}=1$ and all the other elements are 0. The off-diagonal elements remain zero throughout all the evolution, and the diagonal ones evolve as
\begin{equation}
    \left\{
    \begin{array}{cc}
        \rho_{\mu\mu}(t)= \frac{e^{-\gamma t}\left(N-1+e^{\gamma t}\right)}{N} & \mbox{for } \mu=\alpha\\
        \rho_{\mu\mu}(t)= \frac{e^{-\gamma t}\left(-1+e^{\gamma t}\right)}{N} & \mbox{for } \mu \neq \alpha.
        
    \end{array}
    \right.
\end{equation}
So, the unitary transformation optimized to discharge the charged state is the same as the one optimized to discharge the state at any time, since the population profile remains similar: the highest one on the highest eigenvalue and all the others equal to a smaller one. This can be seen in Fig. \ref{fig:hs_energy}, where we plotted the work extracted with the three unitary discharges discussed before for the $N=3$ and $N=4$ cells over time.
This kind of behavior drastically simplifies the optimization of the work extraction, since the optimal discharge operation is the same obtained for a pure initial state. Notice that this is possible only under the assumption that the noise parameter $\gamma$ is uniform over all the noise channels.
\subsection{Stochastic quantum Walks for minimal ring cells}
\label{sec:app_NC_SQW}
\begin{figure}[h!]
    \centering
    \includegraphics[width=0.48\linewidth]{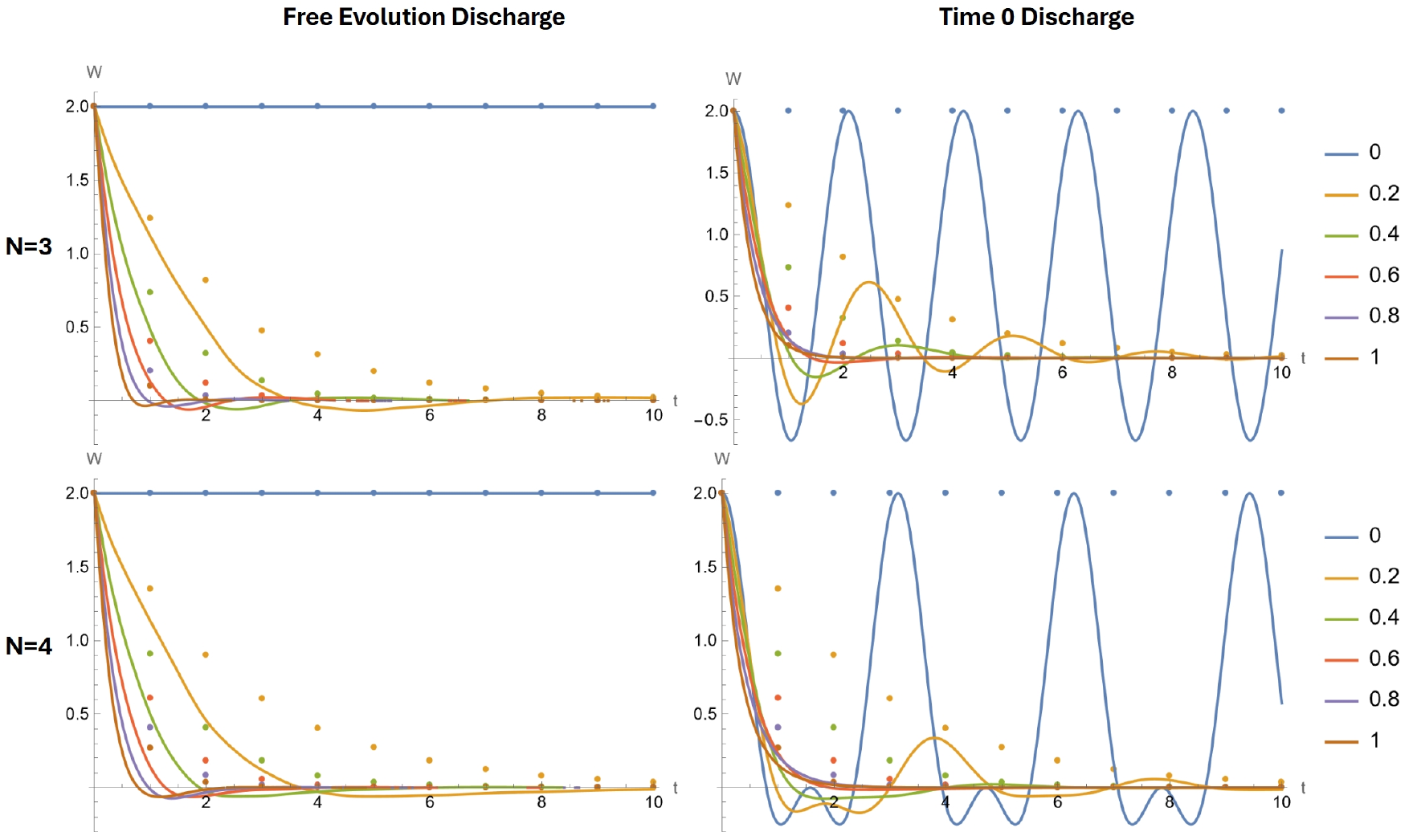}
        \includegraphics[width=0.465 \linewidth]{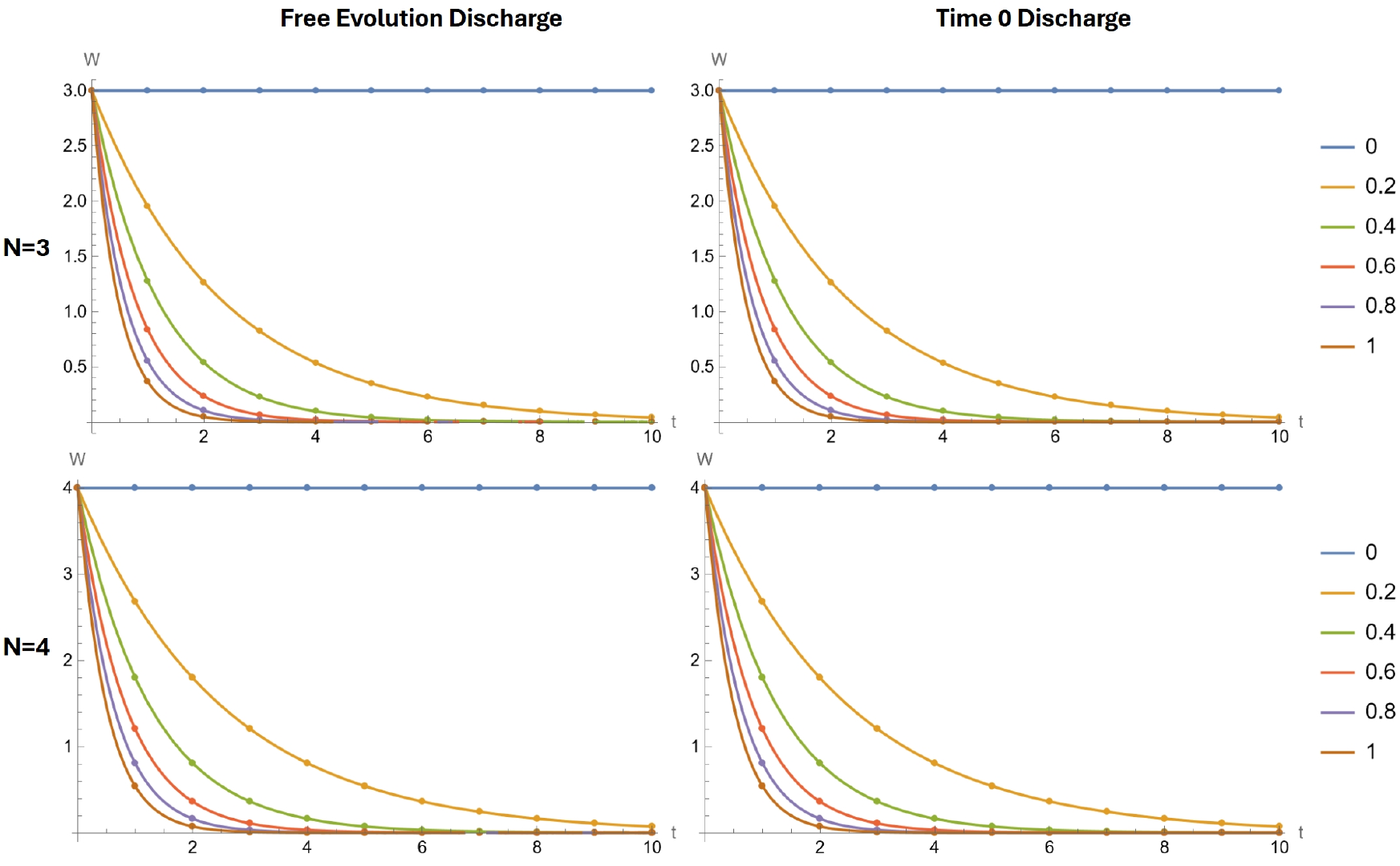}
    \caption{Work extraction under stochastic decoherence for different unitary discharges, number of sites and noise parameters (depicted with different colors). The initial state is chosen as an arbitrary site of the ring (left panels) or the highest energy eigenstate (right panels). 
    The free evolution discharge column represents the work extracted with $\mathcal U_{free}$ while the work extracted with $\mathcal U_0$ is in the second column. In all the plots, the dots represent the Ergotropy, i.e. the the work extracted using $\mathcal U_{erg}$ (note that they all coincide for the highest energy eigenstate).}
    \label{fig:stochastic}
\end{figure}

We put here these results obtained for stochastic quantum walks because of their similarity with those discussed for Haken-Strobl decoherence. We analyzed the work extracted with the unitary discharging processes discussed in Section \ref{sec:NC}, for the triangle and the square ring cells. We considered the charged state initialized either in the highest energy eigenstate or in an arbitrary site of the ring. The results are shown in the two panels of Fig. \ref{fig:stochastic}, respectively.
\section*{Conflict of Interest}
The authors declare that there are no conflicts of interest.
\section*{Data Availability}
No data have been generated. Codes can be accessed upon reasonable request.
\section*{Ethics}
The authors confirm that there are no ethical concerns associated with this study.
\begin{acknowledgments}
This work has been done under the auspices of GNFM-INdAM and is part of Next Generation 
EU project PRIN22-PNRR QWEST (CUP G53D23006270001). It has been also partially supported 
by the no-profit organization {\em Comitato Quantum}, and by MUR and EU through the projects PRIN22 RISQUE (CUP G53D23001110006), NQSTI-Spoke1-BaC  QBETTER (CUP G43C22005120007), 
NQSTI-Spoke2-BaC QMORE (CUP J13C22000680006). 
\end{acknowledgments}
\bibliographystyle{apsrev4-2}
\bibliography{cqbbib.bib}
\end{document}